\documentclass{article}
\usepackage{iclr2027_conference,times}
\usepackage[T1]{fontenc}
\usepackage{graphicx}
\usepackage{booktabs}
\usepackage{amsmath}
\usepackage[hidelinks]{hyperref}
\usepackage{url}
\usepackage{xcolor}
\usepackage[section]{placeins}
\usepackage{caption}
\iclrfinalcopy
\newcommand{\numtable}[1]{%
  {\small\setlength{\tabcolsep}{6pt}\renewcommand{\arraystretch}{1.2}#1}%
}

\graphicspath{{figures/}{figures/appendix/}}

\title{Do AI weather models miss extremes?\\
Evidence from ten months of verification against European
weather stations}

\author{Marvin Vincent Gabler, Roberto Molinaro, Niall Siegenheim, Henry Martin,\\
\textbf{Mark Frey, Niels Poulsen, Philipp Seitz \& Olivier Lam}\\
Jua.ai\\
\texttt{research@jua.ai}
}

\begin{document}
\maketitle
\begin{abstract}
  AI weather models are often reported to underestimate extremes, but most
  evidence concerns deterministic regression models verified against
  reanalysis. We evaluate twelve physical and AI forecast models against
  ECMWF IFS using ten months of European station observations. The
  evaluation covers 10\,m wind, 2\,m temperature, solar radiation, and
  precipitation within regimes defined from a fixed ERA5 1991--2020
  climatology. We find no uniform AI-specific deficit in the tails. Several
  AI models remain more accurate than IFS under extreme conditions, while
  others deteriorate markedly; comparable variation occurs among physical
  models. Every model nevertheless exhibits a common conditional-error
  pattern, overpredicting low observations and underpredicting high
  observations. Attenuation of extreme values therefore does not imply a
  uniform loss of relative skill: tail performance depends on the model,
  variable, and evaluation setting rather than on whether the forecast is
  produced by AI or physical numerical modelling.
  \end{abstract}

\section{Introduction}

Data-driven weather models have achieved strong performance on standard
forecast metrics, often matching or surpassing operational numerical
weather prediction (NWP) models
\citep{bi2023pangu,lam2023graphcast,bouallegue2024rise}. However, many of
these models are trained using mean-squared error and tend to regress
towards the conditional mean, smoothing small-scale variability and
underestimating extremes, as reported for peak winds and record-breaking
events \citep{olivetti2024,bonavita2024limitations,charltonperez2024,
olivetti2024extremes,zhang2026records}.

The available evidence on this behaviour remains nevertheless narrow. Most
existing literature concerns deterministic regression models evaluated
against reanalysis. Probabilistically trained models, generative ensembles,
and high-resolution regional models have received less attention, especially
when verified against weather stations, where errors can be substantially
larger \citep{jin2024weatherreal,gupta2025mausam,wagner2025stationbench}.
Among these, generative models are particularly relevant because they
represent a distribution of possible outcomes rather than a single
conditional estimate, and recent work has shown promising performance for
weather extremes and high-resolution regional forecasting
\citep{price2024gencast,pathak2026stormcast,molinaro2024gencfd}.

A further limitation is that underestimation is often attributed specifically
to AI, while NWP models may also underestimate the same observations.  Hence, the relevant question is not whether AI
models underestimate extremes, but whether they are less accurate than NWP
when extremes occur. An
AI model can underestimate an extreme observation while remaining
closer to it than an NWP forecast.

We test this using twelve forecast models, and compare against ECMWF IFS as the
reference. Forecasts are evaluated over ten months against European
synoptic, solar-radiation, and rain-gauge observations. The analysis covers
10\,m wind, 2\,m temperature, hourly solar radiation, and hourly
precipitation at lead times up to 48\,h. We evaluate extremes in two
complementary ways: first, by stratifying forecasts according to locally
defined climatological regimes; and second, through externally documented
record-breaking temperature events and major European windstorms.

Our main contributions are:

\begin{itemize}

\item We extend station-based evaluation of weather extremes to
a wide range of NWP and AI weather models, including deterministic regression models, CRPS-trained and perturbations-based global ensembles, and
high-resolution regional generative models.

\item We distinguish conditional forecast errors from relative accuracy,
testing whether AI models that underpredict extreme observations are also
less accurate than the NWP reference under the same conditions.

\item We evaluate performance across four surface variables, forecast
leads, countries, and model classes, and complement the climatological
analysis with externally documented temperature-record events, major
windstorms, and heavy-precipitation events.

\end{itemize}
\section{Methods}
\label{sec:data}

We begin by describing the forecast models and station observations, then we introduce the
climatological regimes, bias correction, and evaluation protocol.
Supporting definitions, robustness checks, and complete numerical results
are organised in Appendix~\ref{app:supp}.

\subsection{Models}

Let \(u(t)\) denote the atmospheric state at time \(t\). Weather forecasting
amounts to approximating its evolution from an initial state \(u(t_0)\),
\begin{equation}
\widehat{u}_m(t_0+\tau)
=
\mathcal{F}_m^\tau\bigl(u(t_0)\bigr),
\end{equation}
where \(m\) indexes the forecast model and \(\tau\) is the forecast lead
time. NWP models approximate \(\mathcal{F}_m^\tau\) through numerical
integration of the governing equations on a discrete grid, whereas AI models
learn this mapping from historical atmospheric states. Forecasts may be
deterministic, consisting of a single state, or probabilistic, consisting of
an ensemble,
\[
\left\{\widehat{u}_m^{(k)}(t_0+\tau)\right\}_{k=1}^{K}.
\]

We evaluate twelve forecast models against the high-resolution deterministic
ECMWF IFS forecast, which serves as the reference
(Table~\ref{tab:models}; Appendix~\ref{app:setup}). The comparison spans
NWP and AI models, global and regional domains, and deterministic and
ensemble forecasts.

The NWP models are ECMWF ENS, NOAA GFS, DWD ICON Global, and DWD ICON-EU.
The AI models include ECMWF AIFS \citep{lang2024aifs}, Microsoft Aurora
\citep{bodnar2025aurora}, and Jua EPT-2 Reasoning
\citep{molinaro2025ept2}, all global deterministic regression models. The
global ensembles are ECMWF AIFS ENS \citep{lang2024aifscrps}, a 51-member
ensemble trained with a continuous ranked probability score objective and
learned noise injection, and Jua EPT-2e \citep{molinaro2025ept2}, a
perturbation-based ensemble. The regional models are Jua EPT-2 HRRR
\citep{jua2025hrrr} and EPT-2.1 Europa \citep{jua2026europadocs},
generative ensembles based on the methodology of
\citet{molinaro2026downscaling,molinaro2024gencfd}, and EPT-2.1 Helios
\citep{jua2026helios}, a solar-specialised regional model.

Models are evaluated only for variables with compatible outputs.
Precipitation is available from the NWP models and from EPT-2 Reasoning,
EPT-2e, EPT-2 HRRR, and EPT-2.1 Europa. AIFS, AIFS ENS, and Aurora are
restricted to wind and temperature because their radiation and precipitation
outputs cannot be matched to the hourly verification data.

\subsection{Station observations}
\label{sec:stations}

Forecasts are verified against three complementary in situ datasets:
roughly operational surface observations for 10\,m wind and 2\,m temperature,
500 solar-radiation observations for hourly global horizontal irradiance, and
observations from 4{,}005 rain gauges for hourly precipitation. Specific
contributing networks and data providers are listed in
Appendix~\ref{app:setup}.

\subsection{Observed-value regimes}
\label{sec:thresholds}

Regimes are defined independently of the evaluation period using a fixed
ERA5 1991--2020 climatology \citep{hersbach2020era5}. ERA5 fields are
bilinearly interpolated from the \(0.25^\circ\) grid to each station. For
each station, calendar day, and UTC hour, the reference sample comprises
values from the same UTC hour within a cyclic \(\pm15\)-day window over the
30-year period, yielding 930 values before variable-specific filtering.

Percentile thresholds are computed separately from each reference sample.
Wind and temperature use all reference values, whereas solar radiation uses
positive ERA5 GHI values only. For precipitation, values below
\(0.1\,\mathrm{mm}\) define a separate dry regime, and P50, P75, and P95 are
computed from wet reference hours only. Complete definitions and
meteorological labels are given in Tables~\ref{tab:regimes} and
\ref{tab:precipregimes}. This local-climatology construction follows established practice for climate
extremes \citep{zhang2011indices,kleintank2009guidelines,
perkins2013heatwaves}.

Each verifying observation \(y\) is assigned to a
regime by comparison with the thresholds for its station, calendar day, and
UTC hour.

\subsection{Bias correction}
\label{sec:debias}

Persistent model-wide biases can dominate forecast errors and obscure how
performance changes across observed-value regimes. To reduce their
influence, we apply a causal rolling bias correction before computing skill.
For wind, temperature, and daytime solar radiation, the correction is
additive. For model \(m\), week \(w\), and valid UTC hour \(h\), let
\begin{equation}
\mathcal{D}_{m,w,h}
=
\left\{
\left(f_m(s,t_0,\tau),y(s,t_0+\tau)\right):
\begin{array}{l}
\text{the forecast--observation pair is available},\\
t_0+\tau \text{ lies in the four weeks preceding week } w,\\
\operatorname{hour}_{\mathrm{UTC}}(t_0+\tau)=h
\end{array}
\right\}.
\end{equation}
Thus, \(\mathcal{D}_{m,w,h}\) pools matched forecast--observation pairs
across European stations, initialisations, and leads at the same valid UTC
hour. We estimate the bias as
\begin{equation}
\bar b_{m,w,h}
=
\frac{1}{|\mathcal{D}_{m,w,h}|}
\sum_{(f,y)\in\mathcal{D}_{m,w,h}}(f-y),
\qquad
\widetilde f_m(s,t_0,\tau)
=
f_m(s,t_0,\tau)-\bar b_{m,w,h(t_0+\tau)}.
\end{equation}
For ensemble models, the same shift is applied to every member.

An additive correction is unsuitable for zero-bounded, right-skewed
precipitation. We instead apply a multiplicative correction
\begin{equation}
a_{m,w}
=
\operatorname{clip}_{[0.33,\,3]}
\left(
\frac{\sum_{\mathcal{P}_w}y}
     {\sum_{\mathcal{P}_w}f_m}
\right),
\qquad
\widetilde f_m(t)=a_{m,w}f_m(t),
\end{equation}
where \(\mathcal{P}_w\) contain forecasts and observations from
the preceding four weeks, pooled across initialisations and leads.
This multiplicative correction adjusts systematic amount biases while
preserving zero forecasts. Both procedures use only prior information and do
not alter observation-defined regime membership or force the conditional
error within each regime to zero.

\subsection{Evaluation protocol}
\label{sec:aggregation}

We assess forecast accuracy relative to the high-resolution deterministic
ECMWF IFS forecast. For a given variable, let
\(c=(s,t_0,\tau)\) denote an evaluation cell at station \(s\),
initialisation time \(t_0\), and lead time \(\tau\), verified at
\(t_0+\tau\). For model \(m\), let \(\mathcal{C}_m\) denote the set of cells
for which the bias-corrected forecast, the IFS forecast, and the verifying
observation are all available. Each comparison therefore uses
forecast--observation pairs matched in station, initialisation time, lead
time, and valid time.

For the retained cells \(\mathcal{C}_m\), MAE skill relative to IFS is
\begin{equation}
S_{m,\mathcal{C}_m}
=
100\left(
1-\frac{\operatorname{MAE}_{m,\mathcal{C}_m}}
{\operatorname{MAE}_{\mathrm{IFS},\mathcal{C}_m}}
\right).
\end{equation}
Positive skill indicates lower MAE than IFS, whereas negative skill
indicates higher MAE. Initialisation times are restricted to the 00, 06, 12,
and 18\,UTC cycles.

The evaluation covers thirteen countries: DE, FR, GB, ES, IT, PL, NL, BE,
CH, NO, AT, CZ, and DK, and it is organised into two tracks. Track~A covers
1 September 2025 to 30 June 2026. Wind and temperature are evaluated at
6-hourly leads from 6 to 48\,h, whereas solar radiation and hourly
precipitation are evaluated at all leads from 1 to 48\,h. Track~B covers
1 March to 30 June 2026 and compares EPT-2.1 Europa, EPT-2 HRRR, and
ICON-EU on a common hourly schedule at leads from 1 to 48\,h; EPT-2.1 Helios
is additionally included for solar radiation. Italy and Poland are excluded
from precipitation because of missing observations.

Forecasts are bilinearly interpolated to station locations
\citep{molinaro2026downscaling}. Ensemble models are evaluated through their
means, so the analysis assesses point-forecast accuracy rather than
probabilistic calibration or ensemble spread.

It should be observed that none of the models was trained or fine-tuned on the verification
observations. Historical station errors enter only through the causal
bias-correction procedure described above. Uncertainty in aggregated skill
is quantified using a leave-one-month-out jackknife, with error bars
representing one standard error.

\section{Results}

\subsection{Skill across the observed distribution}
\label{sec:regimes}

\begin{figure}[!hbp]
\centering
\includegraphics[width=0.88\textwidth]{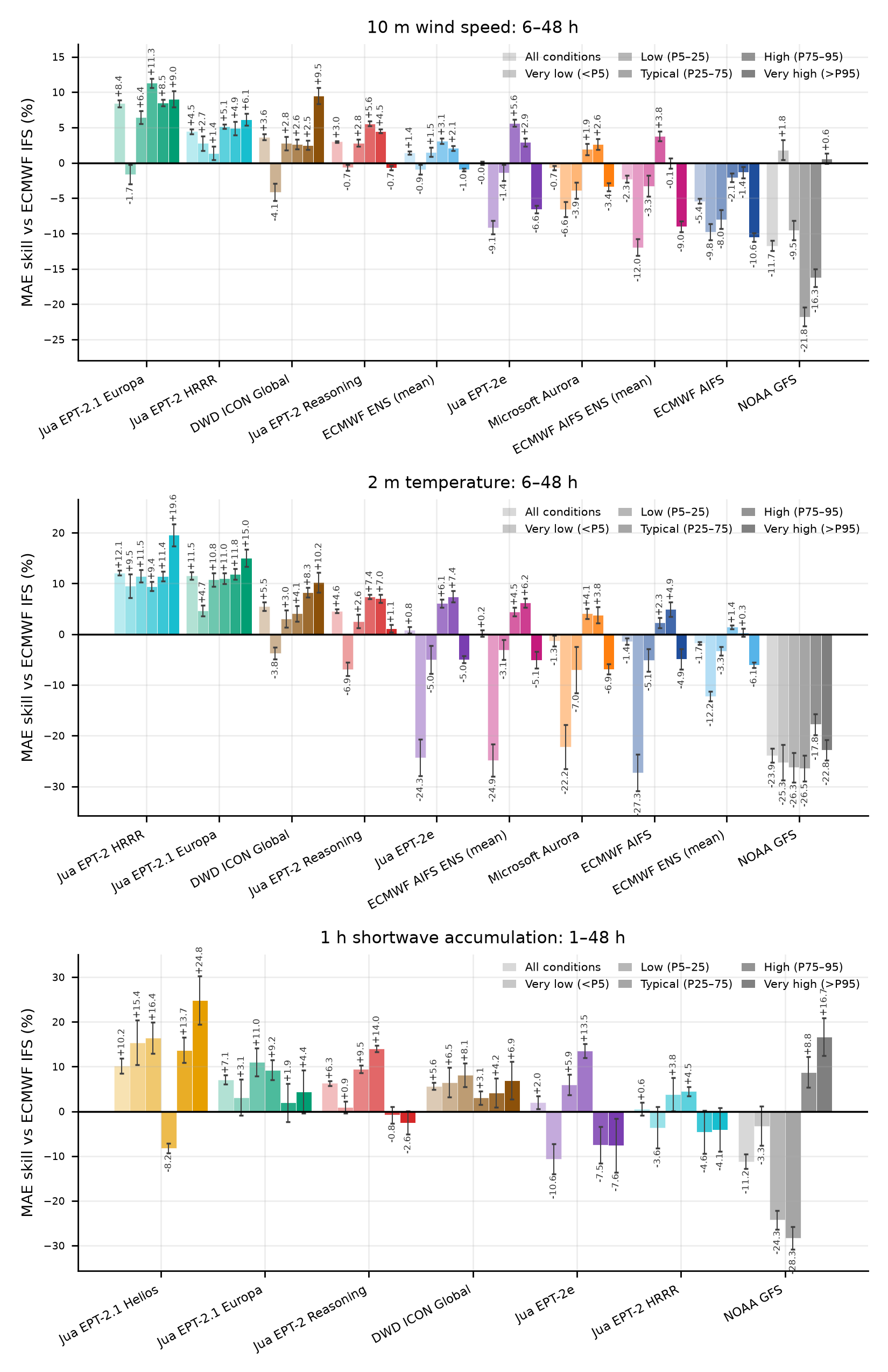}
\caption{Track~A MAE skill relative to ECMWF IFS by observed-value regime,
with \(\pm1\) jackknife SE. Wind and temperature use 6--48\,h leads;
solar radiation uses 1--48\,h.}
\label{fig:bars}
\end{figure}

\begin{figure}[!hbp]
\centering
\includegraphics[width=0.9\textwidth]{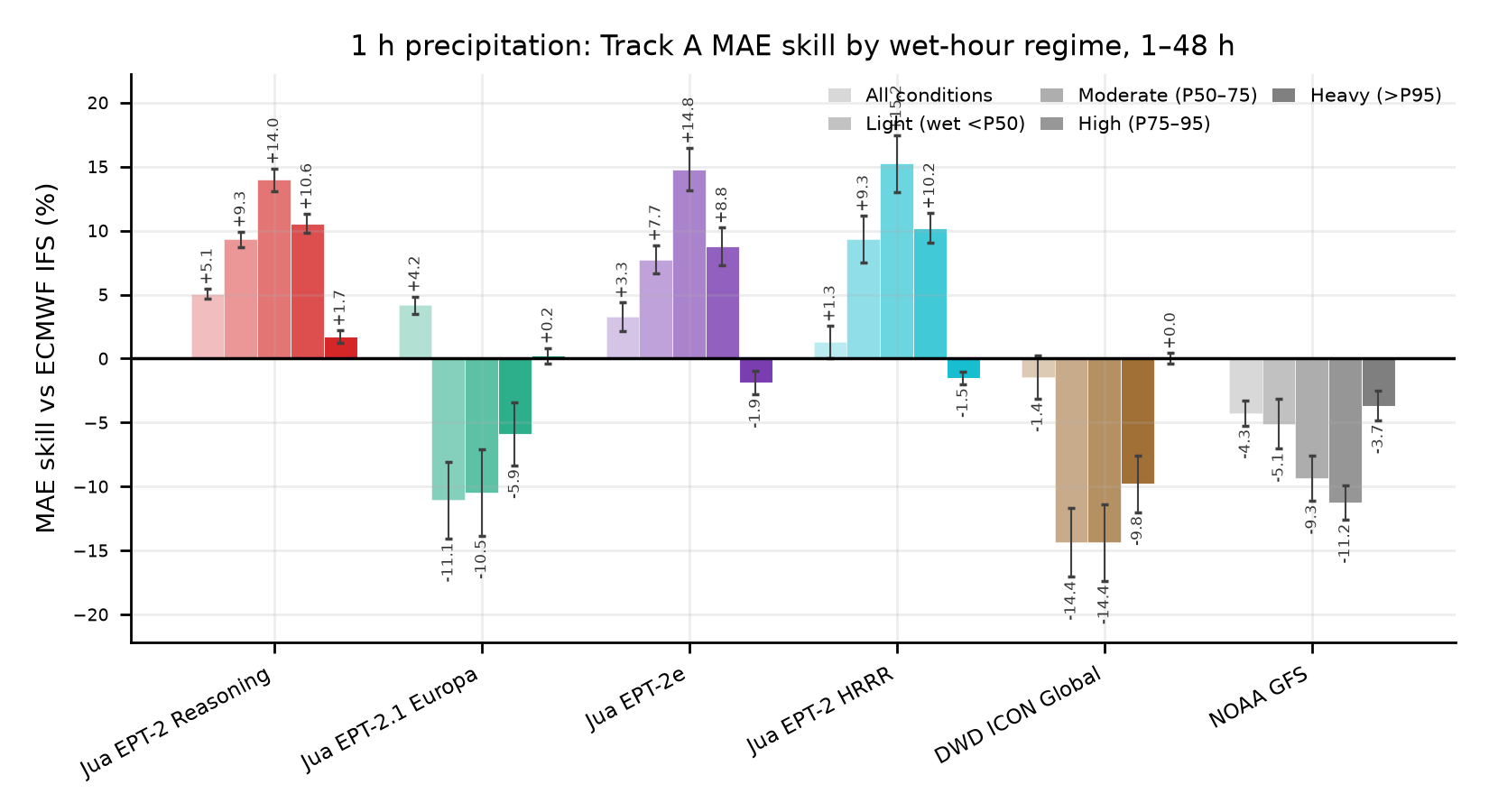}
\caption{Track~A MAE skill relative to ECMWF IFS for hourly precipitation
by observed-value regime over 1--48\,h, with \(\pm1\) jackknife SE.}
\label{fig:precipbars}
\end{figure}

Track~A results are reported in Figures~\ref{fig:bars} and
\ref{fig:precipbars}. Figure~\ref{fig:bars} shows MAE skill relative to IFS
by observed-value regime for wind, temperature, and solar radiation, while
Figure~\ref{fig:precipbars} shows the corresponding precipitation results.

Tail skill does not follow a uniform AI--physical ordering. EPT-2.1 Europa
leads the wind comparison overall and remains competitive with ICON Global
in the gale-force regime. EPT-2 HRRR and EPT-2.1 Europa perform most strongly
for temperature, including in the heat regime. By contrast, AIFS and AIFS ENS
remain below IFS in the wind and temperature tails despite the ensemble's
improved performance near the centre of the distribution. The largest heat
deficit belongs to the NWP model GFS.

Solar radiation follows a different pattern. EPT-2.1 Helios leads overall
and under overcast and near-clear-sky conditions but falls below IFS under
typical irradiance, where EPT-2 Reasoning performs most strongly. Its
advantage is therefore concentrated in the tails rather than distributed
uniformly across conditions.

For precipitation, the clearest gains occur in the middle of the wet-hour
distribution (Figure~\ref{fig:precipbars}). EPT-2 HRRR, EPT-2e, and
EPT-2 Reasoning improve on IFS from moderate through high rain, whereas the
physical global models remain below the reference. These gains narrow at the
heaviest intensities. Europa is the exception: its positive all-conditions
score comes from dry hours, while it is worse than IFS in every wet-hour
regime.

We additionally evaluate detection of heavy-rain events using the local
wet-hour P95 threshold. For each forecast--observation pair, the observation
and forecast are each labelled a heavy-rain event if their hourly
accumulation exceeds the same local wet-hour P95 threshold. All evaluated
precipitation models forecast fewer P95 exceedances than observed. Europa
and ICON Global nevertheless achieve a slightly higher critical success
index than IFS (Figure~\ref{fig:precipcat};
Table~\ref{tab:precipcat}).

We further assess the robustness of the Track~A results to forecast lead,
the inclusion of short-range forecasts, and country-level aggregation. The
lead-time and 1--12\,h analyses preserve the principal model ordering
(Appendix~\ref{app:tracka}). Country-level results also show that the main
temperature and wind patterns occur across multiple countries rather than
being dominated by one country (Appendix~\ref{app:country}).

\subsection{Shared conditional-error pattern}
\label{sec:fingerprint}

\begin{figure}[!hbp]
\centering
\includegraphics[width=0.9\textwidth]{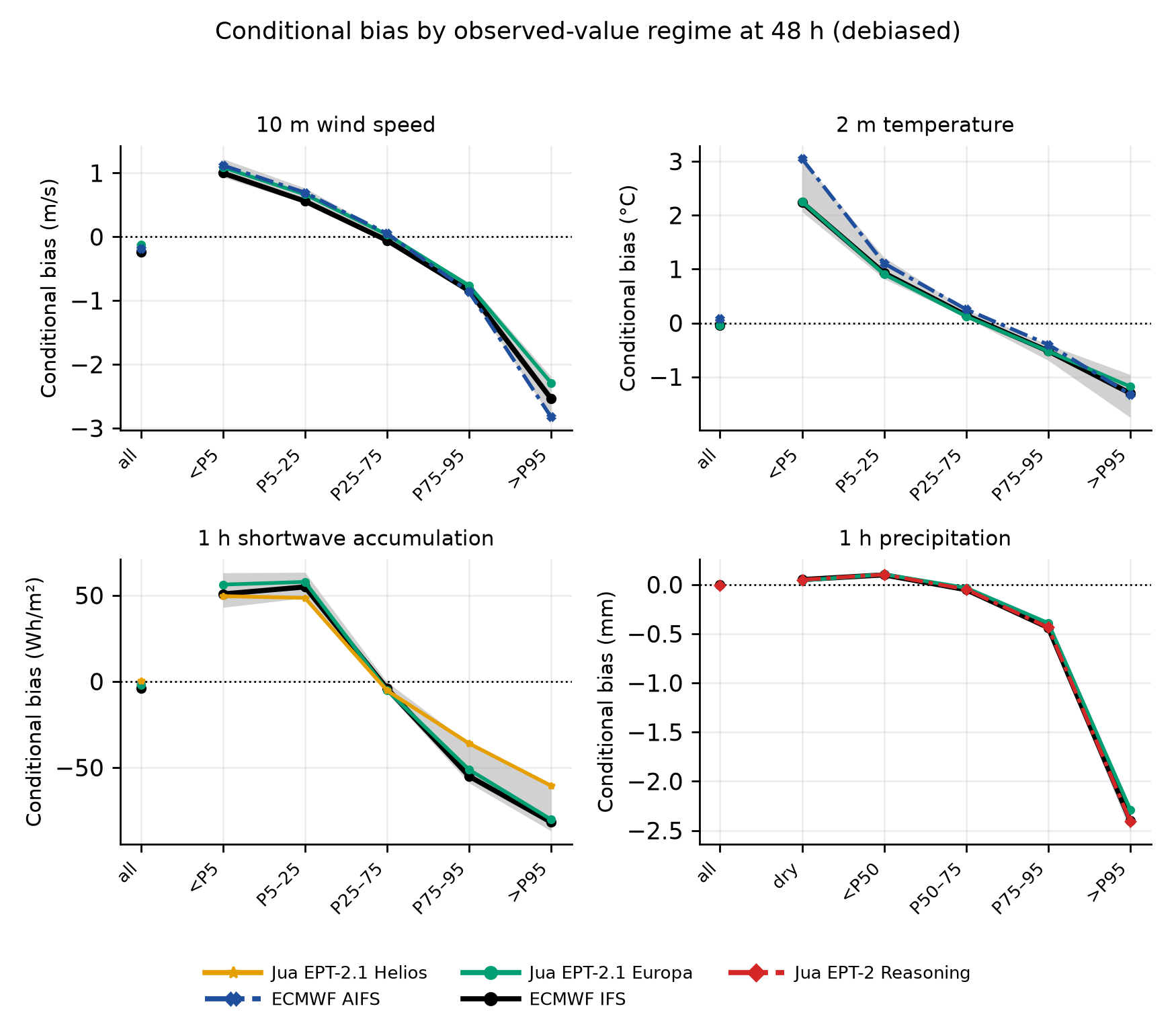}
\caption{Mean forecast-minus-observation error at 48\,h, conditioned on
the observed-value regime after bias correction. The grey envelope spans
all models; selected models are highlighted.}
\label{fig:fingerprint}
\end{figure}

Figure~\ref{fig:fingerprint} reports conditional bias at 48\,h by
observed-value regime. Across all four variables, AI and NWP models,
including IFS, show the same sign pattern: bias is positive for low
observations and negative for high observations. The shift across regimes is
larger than the differences between models at a fixed regime. This provides
further evidence that conditional-error behaviour does not differ
systematically between AI and NWP models, despite differences in relative
MAE.

\subsection{Record-breaking temperature events and major windstorms}
\label{sec:events}

To complement the climatological-regime analysis, we evaluate five externally
documented record-breaking temperature events and seven major European
windstorms. Event windows and affected-country footprints were specified from
external reports before scoring. A forecast sample is retained only when IFS
and every Track~A model included in the comparison share the same station,
valid time, initialisation, and 6-hourly lead from 6 to 48\,h. Each model
therefore uses identical event windows and forecast--observation samples.
Full event definitions, provenance, scoring details, and numerical results
are reported in Appendix~\ref{app:officialevents}
(Tables~\ref{tab:officialeventcatalog} and
\ref{tab:officialeventskill}).

\begin{figure}[tbp]
\centering
\includegraphics[width=\textwidth]{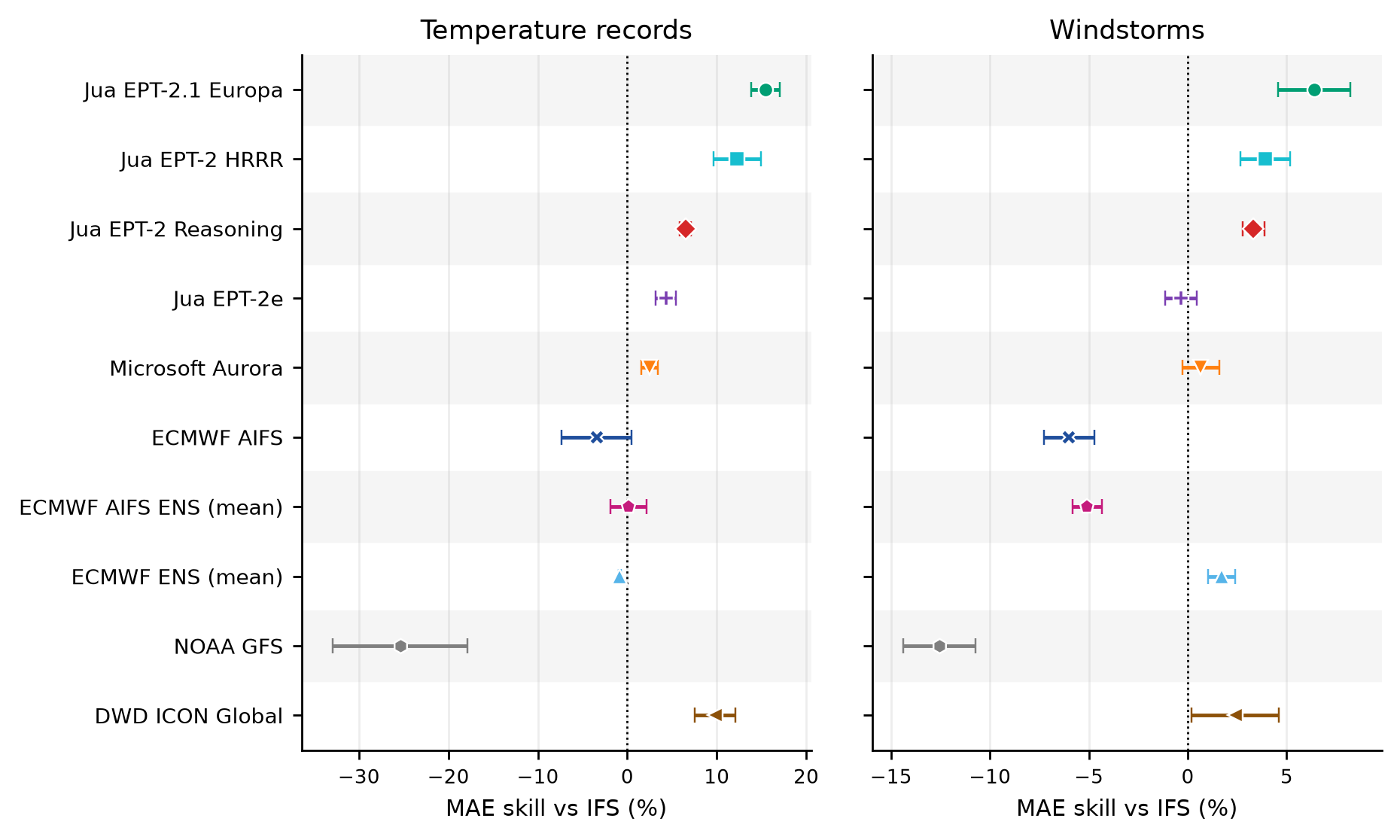}
\caption{MAE skill relative to ECMWF IFS during five documented
record-breaking temperature events (left; \(n=231{,}065\)) and seven major
European windstorms (right; \(n=123{,}847\)), aggregated over common
station/valid-time/lead samples from 6--48\,h after bias correction. Error
bars show one leave-one-event-out jackknife SE.}
\label{fig:officialevents}
\end{figure}

Figure~\ref{fig:officialevents} reports MAE skill relative to IFS during the
documented windstorms and temperature events. Europa has the largest positive
skill in both sets of events. During the temperature events, performance
varies substantially across models, with GFS showing the largest deficit.
During the windstorms, most models remain close to IFS, whereas AIFS, AIFS
ENS, and GFS remain below the reference. These source-defined cases therefore
support the model-specific pattern found in the climatological-regime
analysis, rather than a uniform AI-specific loss of skill during extremes.

\subsection{Regional comparison}

\begin{figure}[!hbp]
\centering
\includegraphics[width=0.88\textwidth]{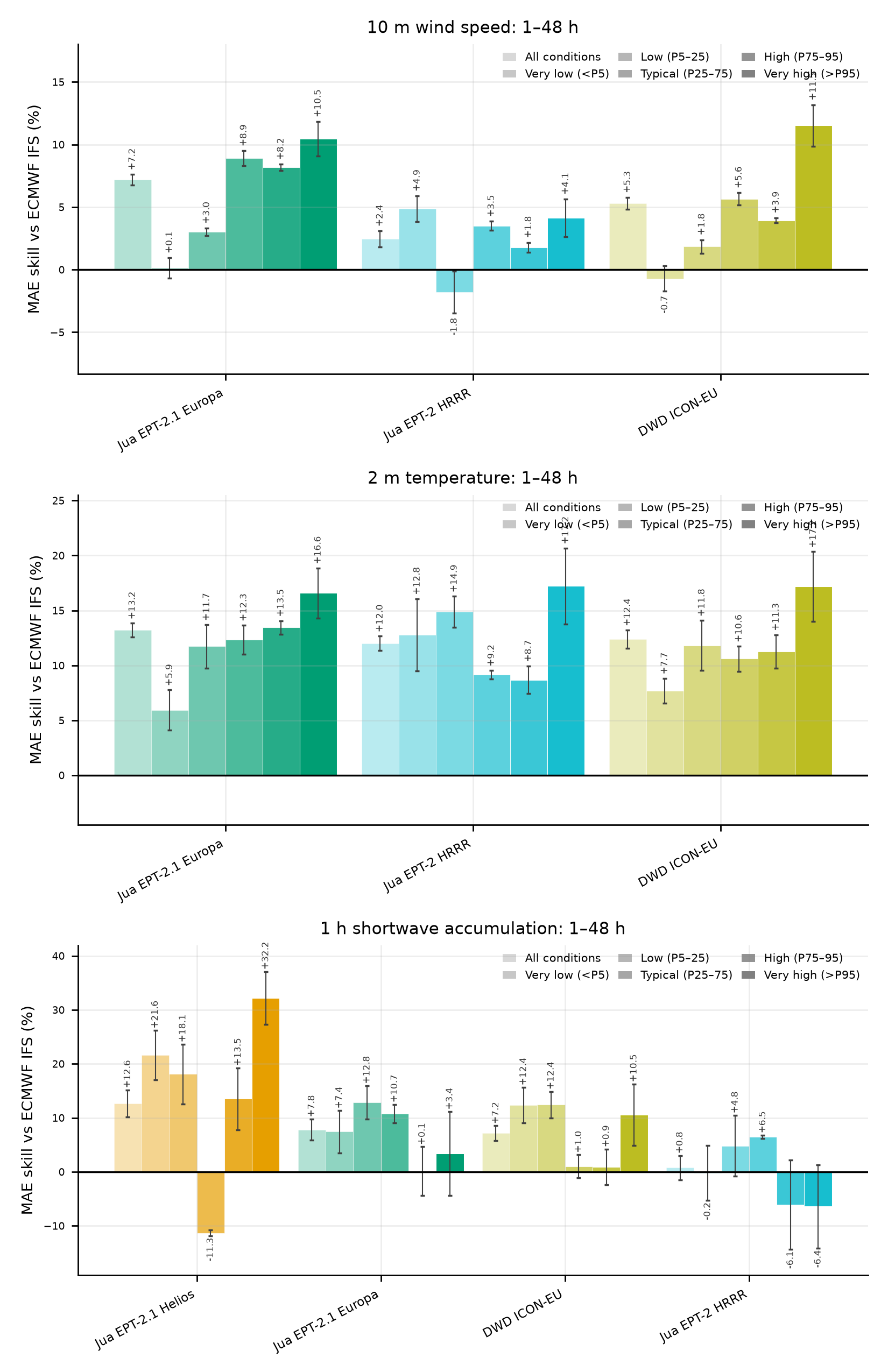}
\caption{Track~B MAE skill relative to ECMWF IFS by observed-value regime,
aggregated over 1--48\,h from March to June 2026, with \(\pm1\)
jackknife SE. Wind and temperature compare EPT-2.1 Europa, EPT-2 HRRR,
and ICON-EU; the solar panel additionally includes EPT-2.1 Helios.
Models are ordered by all-conditions skill within each panel.}
\label{fig:trackbbars}
\end{figure}

Finally, we use Track~B to compare EPT-2.1 Europa and EPT-2 HRRR directly
with ICON-EU over their shared European domain from March to June 2026.
Figure~\ref{fig:trackbbars} reports MAE skill relative to IFS by
observed-value regime. Europa retains the strongest overall wind estimate,
while ICON-EU is competitive with the leading AI models in the wind and
temperature tails. EPT-2.1 Helios has the clearest solar advantage.

The supplementary Track~B panels resolve these comparisons by lead time and
over the 1--12\,h range; neither changes this interpretation
(Appendix~\ref{app:trackb}). Complete numerical values are reported in
Appendix~\ref{app:numerical}.

\section{Discussion}

The results distinguish between two questions: whether a model
underestimates an extreme observation, and whether it is less accurate than
IFS when that observation occurs.

Across deterministic, probabilistic, and generative AI models, as well as
NWP models, low observations are generally overpredicted and high
observations are underpredicted. This shared conditional-error pattern does
not yield a consistent class-level ordering. Instead, differences emerge at
the level of individual systems: AIFS and AIFS ENS perform worse than IFS in
the wind and temperature tails, whereas EPT-2 HRRR and EPT-2.1 Europa remain
competitive in the gale-force and heat regimes. The NWP results likewise
vary, with GFS showing the largest heat deficit while ICON-EU is competitive
with the leading regional AI models in the matched Track~B comparison. The
source-defined temperature events and windstorms show the same
model-specific pattern on common event samples: the leading regional systems
remain ahead of IFS, whereas the two AIFS variants and GFS lose skill in at
least one event set.

Relative accuracy also depends on the observed regime and variable. For
example, EPT-2.1 Helios is more accurate than IFS in both solar tails but
less accurate under typical irradiance. Similarly, EPT-2 HRRR, EPT-2e, and
EPT-2 Reasoning improve on IFS for moderate to high precipitation, although
their advantage narrows for the heaviest events.

Within this scope, the results do not support a general weakness of AI
weather forecasting at extremes.

\subsection{Limitations}

The comparisons identify differences in performance, but not their causes.
The systems differ in architecture, training objective, resolution,
training data, initialisation, cadence, and domain
\citep{bonavita2024limitations,subich2025doublepenalty}; consequently,
their performance cannot be attributed independently to AI versus NWP,
regression versus generative training, or global versus regional
resolution. Moreover, evaluating ensembles through their means assesses
point-forecast accuracy, rather than calibration, spread, or event
probabilities.

\subsubsection*{Ethics statement}
\ificlrfinal
All authors are affiliated with Jua.ai AG, which develops several of the
forecast systems evaluated in this study.
\else
The authors have a competing interest with respect to some of the evaluated
forecast systems; details are withheld for anonymity and will be disclosed in
the final version.
\fi
The verification protocol is applied identically across AI and physical
systems, every comparison is reported against the independent ECMWF IFS
reference, and the aggregates and reproduction code described below are
provided to support independent scrutiny of the comparisons. The study uses forecast output and meteorological station
observations only; it involves no human subjects or personal data, and no
station identities or per-station series are released.

\subsubsection*{Reproducibility statement}
Metrics were computed through a content-addressed station-verification API;
requests and responses are archived. The reproduction package includes the
configuration files, the derived aggregates behind every reported number, and
the extraction, scoring, aggregation and figure code. All score-based figures
and tables regenerate offline from those aggregates; station-level
observations and raw forecast fields are not redistributed.
\ificlrfinal
The package is at \url{https://github.com/juaAI/aiweather-extremes}, and
the forecast fields used in the evaluation are available via the Jua
platform (\url{https://jua.ai}).
\else
The package is provided as anonymised supplementary material.
\fi

\subsubsection*{AI use statement}

Generative AI tools were used to review and refine the experimental
methodology; to implement extraction, quality-control, scoring, aggregation,
and figure code, including the cleaning and reformatting of observation and
forecast data; to inspect and interpret generated results; to
search, summarise, and verify related literature and format references; and
to draft, structure, and edit manuscript text. They were not used to generate
synthetic data, to alter individual observation or forecast values, or to
prove mathematical claims. The authors reviewed and tested all AI-assisted code, verified the
reported numbers against the aggregate data, and take full responsibility
for the final content.

\bibliography{references}
\bibliographystyle{iclr2027_conference}

\clearpage
\appendix
\section{Supplementary material}\label{app:supp}
\setcounter{table}{0}
\setcounter{figure}{0}
\renewcommand{\thetable}{S\arabic{table}}
\renewcommand{\thefigure}{S\arabic{figure}}

This supplement provides the model and regime definitions omitted
from the main text, lead-time and short-range robustness checks for Track~A, the
matched regional comparison in Track~B, complete numerical tables, and
country-level regime profiles.

\subsection{Evaluation setup}
\label{app:setup}

Surface wind and temperature observations were obtained through Synoptic
Data from selected operational observing networks, including SYNOP, WIS2,
ASOS/AWOS, and METAR. Solar-radiation observations were compiled from
ARPALomb, BSRN, DMI, DWD, FMI, Frost, GeoSphere Austria, KNMI,
Météo-France, MeteoSwiss, RMI, and SMHI.

Table~\ref{tab:models} identifies the forecasting systems and their broad
model classes. Tables~\ref{tab:regimes} and \ref{tab:precipregimes} give the
observed-value intervals and meteorological labels used throughout the
evaluation.


\begin{table}[htbp]
\centering
\caption{Forecast models included in the evaluation. ECMWF IFS is the
reference model.}
\label{tab:models}
\numtable{\begin{tabular}{llll}
\toprule
Model & Model class & Forecast type & Domain \\
\midrule
ECMWF IFS (reference) & Physical & Deterministic & Global \\
ECMWF ENS & Physical & Ensemble & Global \\
NOAA GFS & Physical & Deterministic & Global \\
DWD ICON Global & Physical & Deterministic & Global \\
DWD ICON-EU & Physical & Deterministic & Regional \\
\addlinespace
ECMWF AIFS & Regression AI & Deterministic & Global \\
Microsoft Aurora & Regression AI & Deterministic & Global \\
Jua EPT-2 Reasoning & Regression AI & Deterministic & Global \\
Jua EPT-2e & Regression AI & Ensemble & Global \\
Jua EPT-2.1 Helios & Regression AI (solar) & Deterministic & Regional \\
\addlinespace
ECMWF AIFS ENS & Probabilistic AI & Ensemble & Global \\
Jua EPT-2 HRRR & Generative AI & Ensemble & Regional \\
Jua EPT-2.1 Europa & Generative AI & Ensemble & Regional \\
\bottomrule
\end{tabular}}
\end{table}

\begin{table}[htbp]
\centering
\caption{Observed-value regimes for wind, temperature, and solar radiation.
For solar radiation, the percentiles refer to positive daytime GHI.}
\label{tab:regimes}
\numtable{\begin{tabular}{lllll}
\toprule
Regime & Definition & Wind & Temperature & Solar \\
\midrule
All       & --                          & all conditions  & all conditions & all daytime \\
Very low  & \(y<\mathrm{P5}\)                         & calm            & severe cold    & heavily overcast \\
Low       & \(\mathrm{P5}\leq y<\mathrm{P25}\)        & light air       & cold           & overcast \\
Typical   & \(\mathrm{P25}\leq y<\mathrm{P75}\)       & moderate        & moderate       & mixed sky \\
High      & \(\mathrm{P75}\leq y<\mathrm{P95}\)       & fresh to strong & warm           & mostly sunny \\
Very high & \(y\geq\mathrm{P95}\)                     & gale force      & heat           & near clear sky \\
\bottomrule
\end{tabular}}
\end{table}

\begin{table}[htbp]
\centering
\caption{Observed-value regimes for hourly precipitation. Wet-hour
percentiles are conditional on ERA5 accumulations of at least \(0.1\)~mm.}
\label{tab:precipregimes}
\numtable{\begin{tabular}{lll}
\toprule
Regime & Definition & Interpretation \\
\midrule
All      & --                                                       & all hours \\
Dry      & \(y<0.1\,\mathrm{mm}\)                                   & dry hour \\
Light    & \(0.1\,\mathrm{mm}\leq y<\mathrm{P50}\)        & light rain \\
Moderate & \(\mathrm{P50}\leq y<\mathrm{P75}\)            & moderate rain \\
High     & \(\mathrm{P75}\leq y<\mathrm{P95}\)            & heavy rain \\
Heavy    & \(y\geq\mathrm{P95}\)                          & most intense hours \\
\bottomrule
\end{tabular}}
\end{table}

\FloatBarrier
\subsection{Track A lead-time, short-range, and precipitation robustness}
\label{app:tracka}

\begin{figure}[htbp]
\centering
\includegraphics[width=\textwidth]{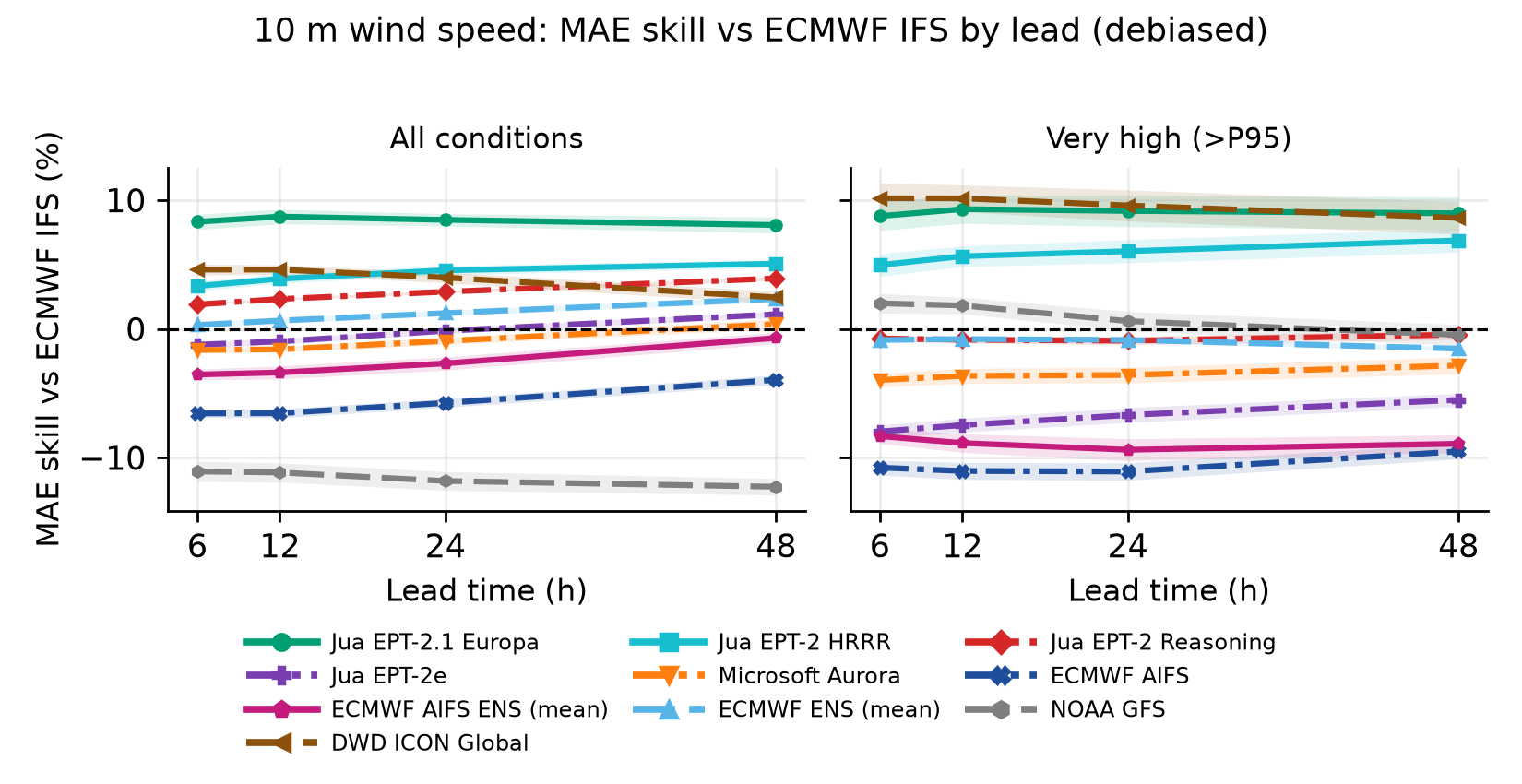}
\caption{Track A MAE skill relative to ECMWF IFS by lead time for 10\,m wind.
The left panel shows all conditions; the right panel shows the gale-force
regime, \(y\geq\mathrm{P95}\). Shading denotes \(\pm1\) jackknife SE.}
\label{fig:leadwind}
\end{figure}

\begin{figure}[htbp]
\centering
\includegraphics[width=\textwidth]{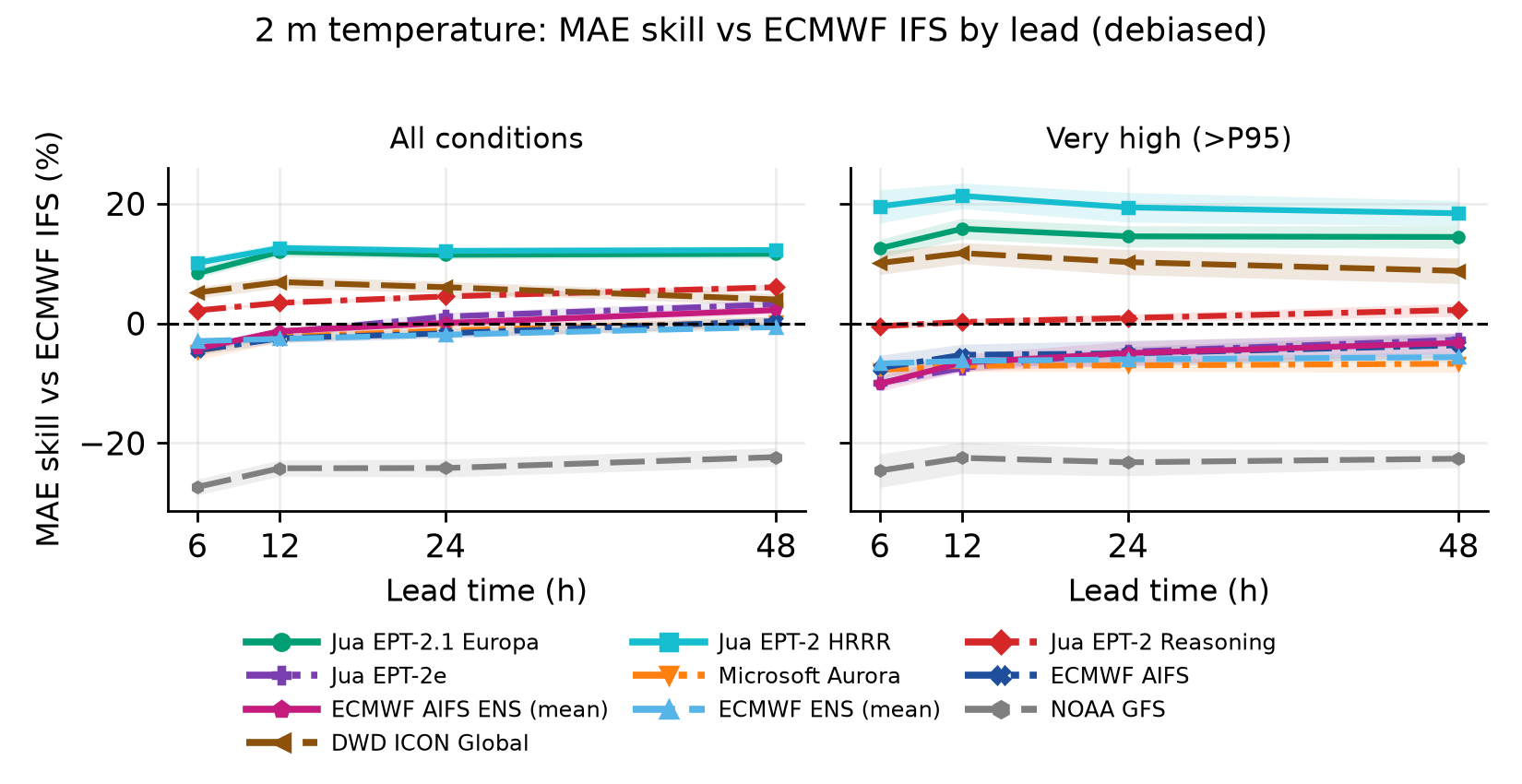}
\caption{Track A MAE skill relative to ECMWF IFS by lead time for 2\,m temperature.
The left panel shows all conditions; the right panel shows the heat
regime, \(y\geq\mathrm{P95}\). Shading denotes \(\pm1\) jackknife SE.}
\label{fig:leadtemp}
\end{figure}

\begin{figure}[htbp]
\centering
\includegraphics[width=\textwidth]{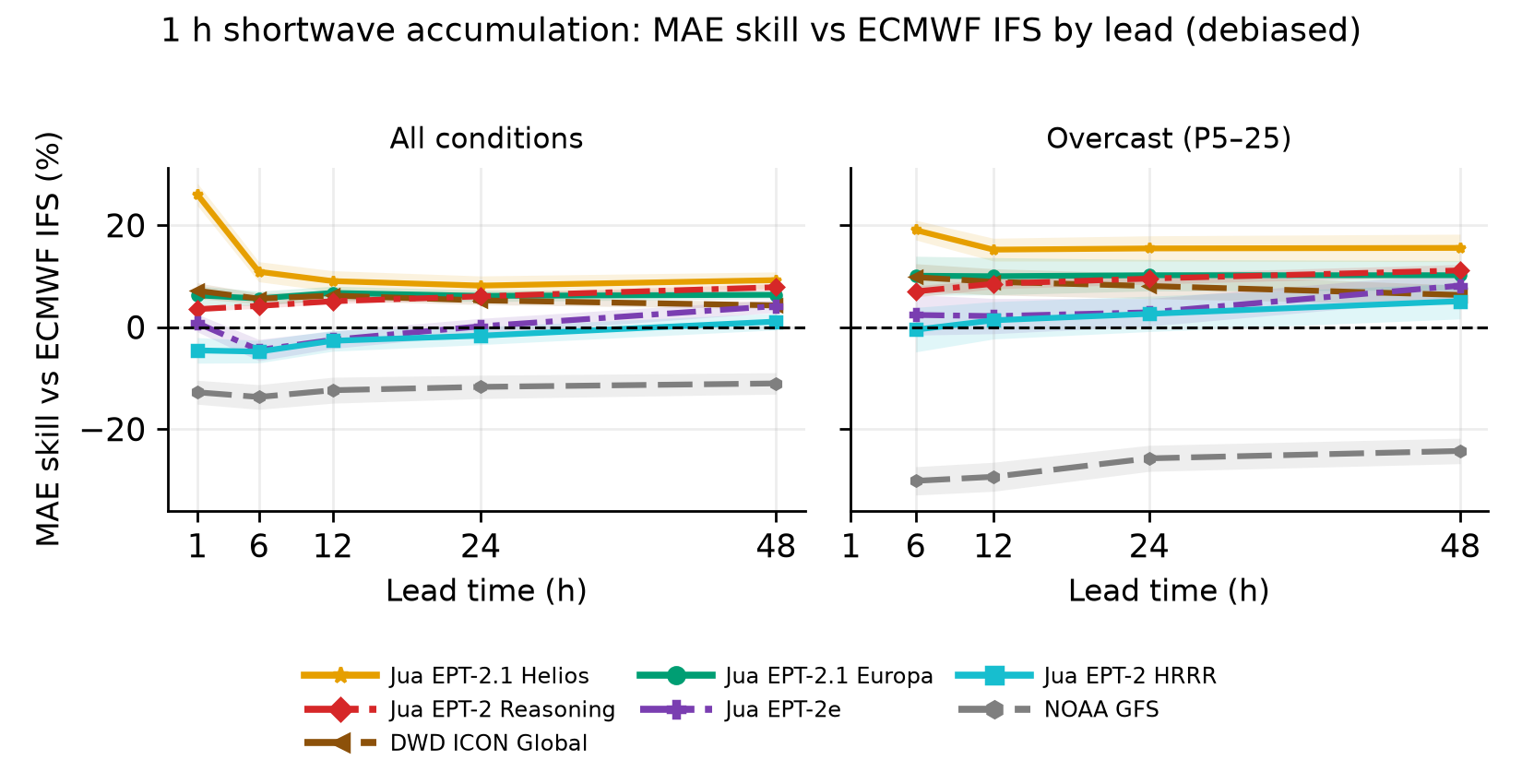}
\caption{Track~A MAE skill relative to ECMWF IFS by lead time for hourly
solar radiation. The left panel shows all daytime conditions; the right
panel shows the overcast regime,
\(\mathrm{P5}\leq y<\mathrm{P25}\). AIFS, AIFS ENS, Aurora, and ECMWF ENS are
omitted because they lack compatible hourly solar output. Shading denotes
\(\pm1\) jackknife SE.}
\label{fig:leadsolar}
\end{figure}

\begin{figure}[htbp]
\centering
\includegraphics[width=\textwidth]{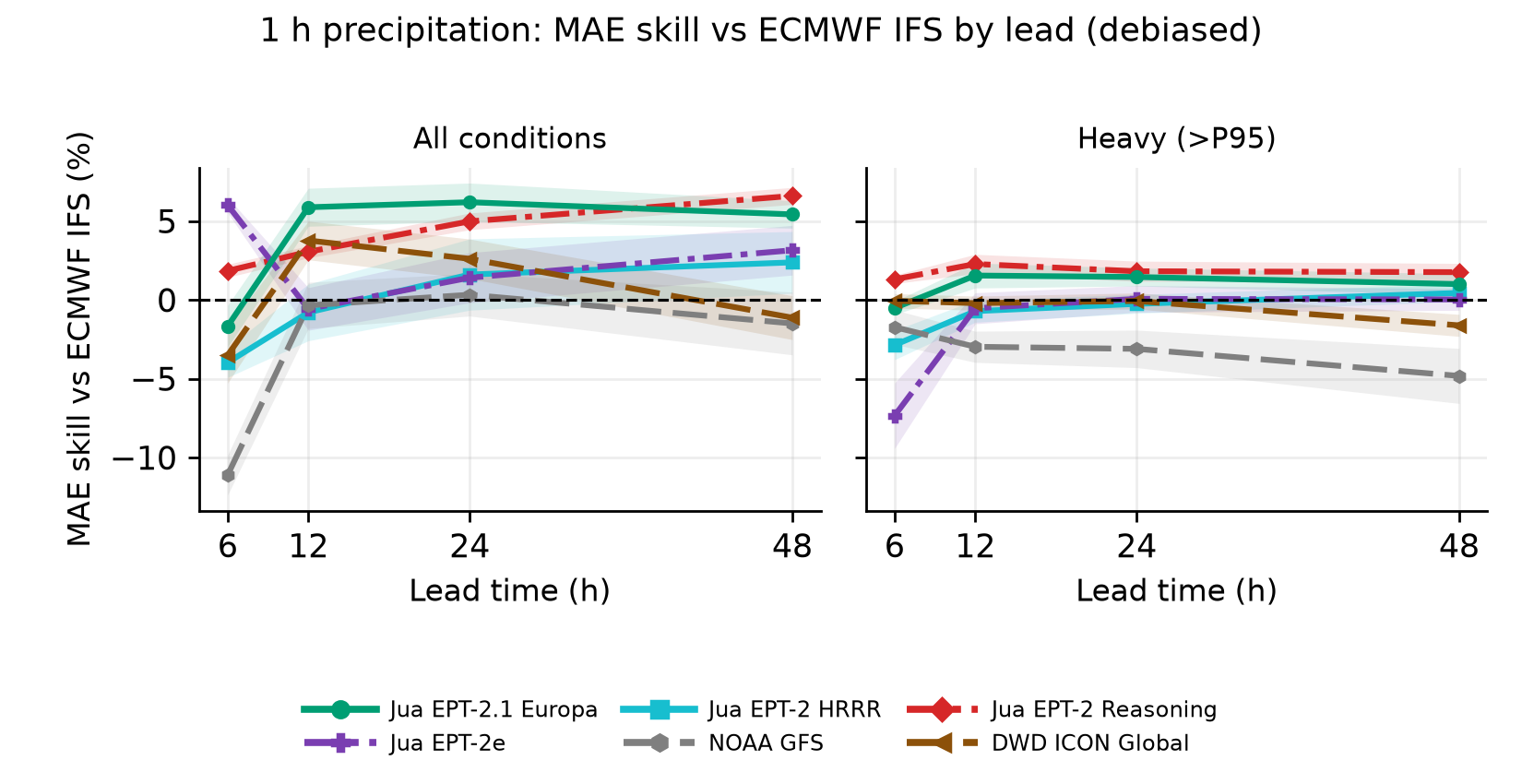}
\caption{Track~A MAE skill relative to ECMWF IFS by lead time for hourly
precipitation. The left panel shows all conditions; the right panel shows
the heavy regime, \(y\geq\mathrm{P95}\). Shading denotes
\(\pm1\) jackknife SE.}
\label{fig:preciplead}
\end{figure}

\begin{figure}[htbp]
\centering
\includegraphics[width=0.88\textwidth]{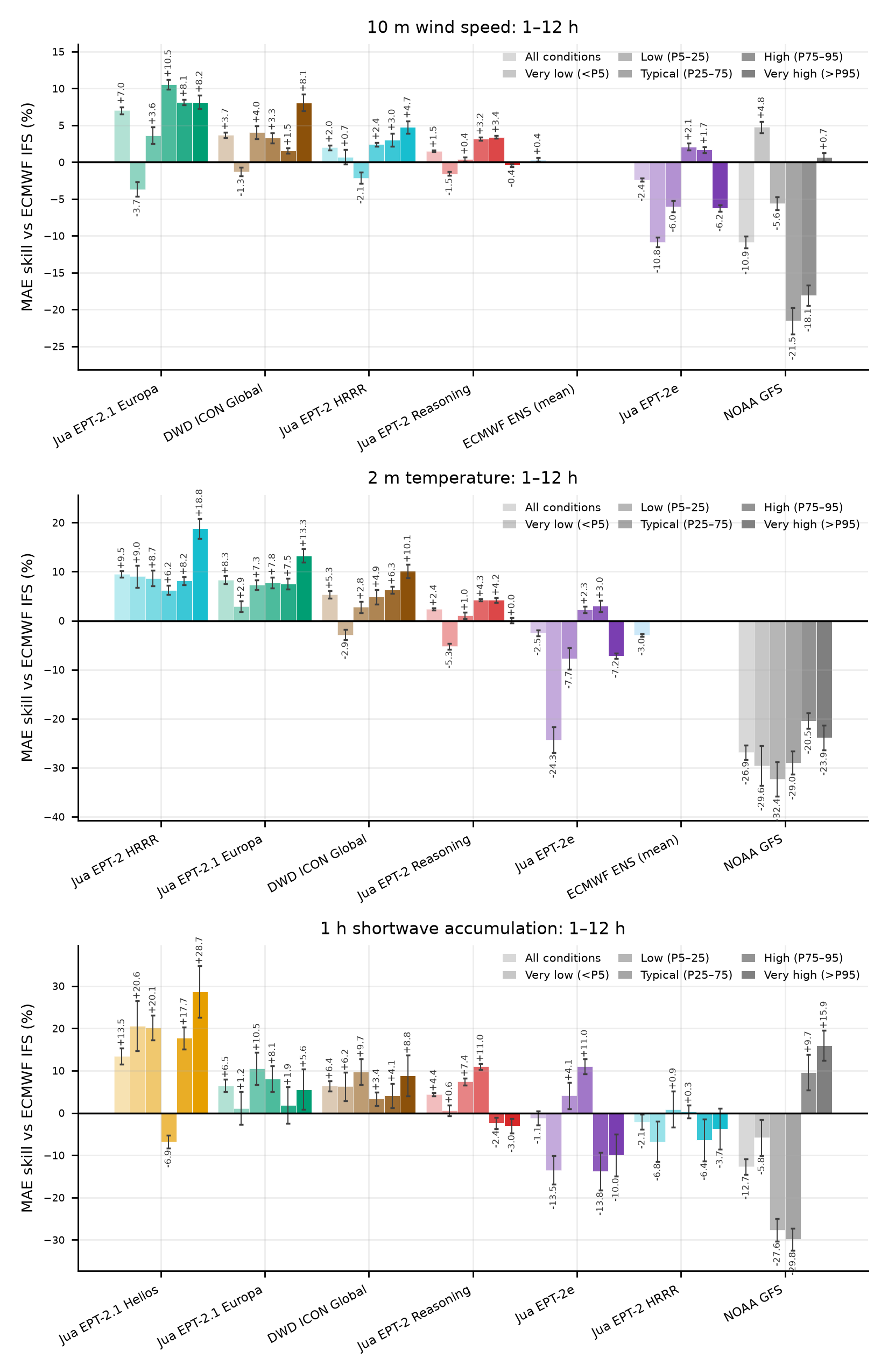}
\caption{Track~A MAE skill relative to ECMWF IFS by observed-value regime
over 1--12\,h, with \(\pm1\) jackknife SE. Wind and temperature are
restricted to models with hourly output; solar radiation includes
EPT-2.1 Helios.}
\label{fig:bars12}
\end{figure}

\begin{figure}[htbp]
\centering
\includegraphics[width=\textwidth]{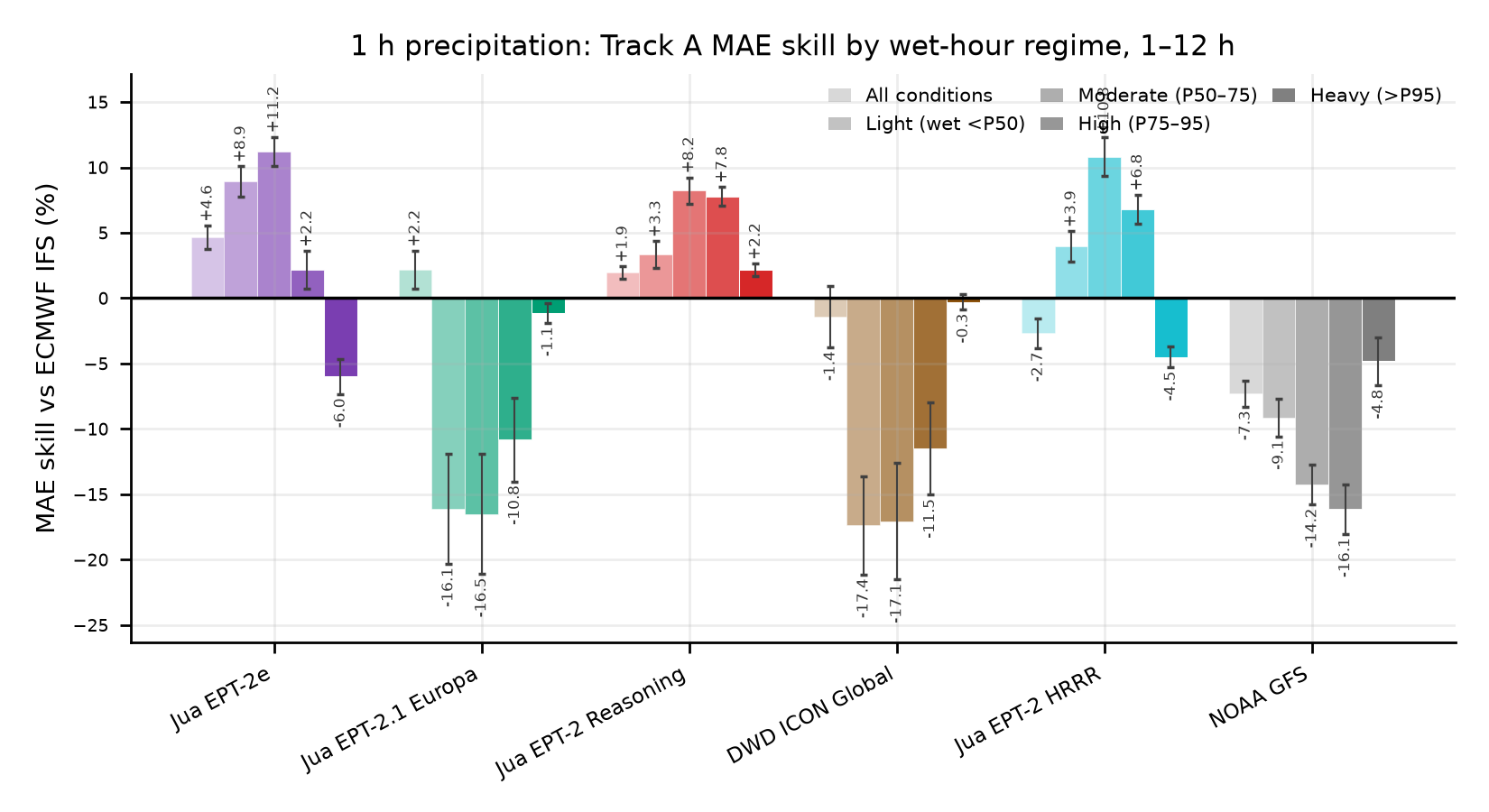}
\caption{Track~A MAE skill relative to ECMWF IFS for precipitation by observed-value regime
over 1--12\,h, with \(\pm1\) jackknife SE }
\label{fig:precipbars12}
\end{figure}

\begin{figure}[htbp]
\centering
\includegraphics[width=\textwidth]{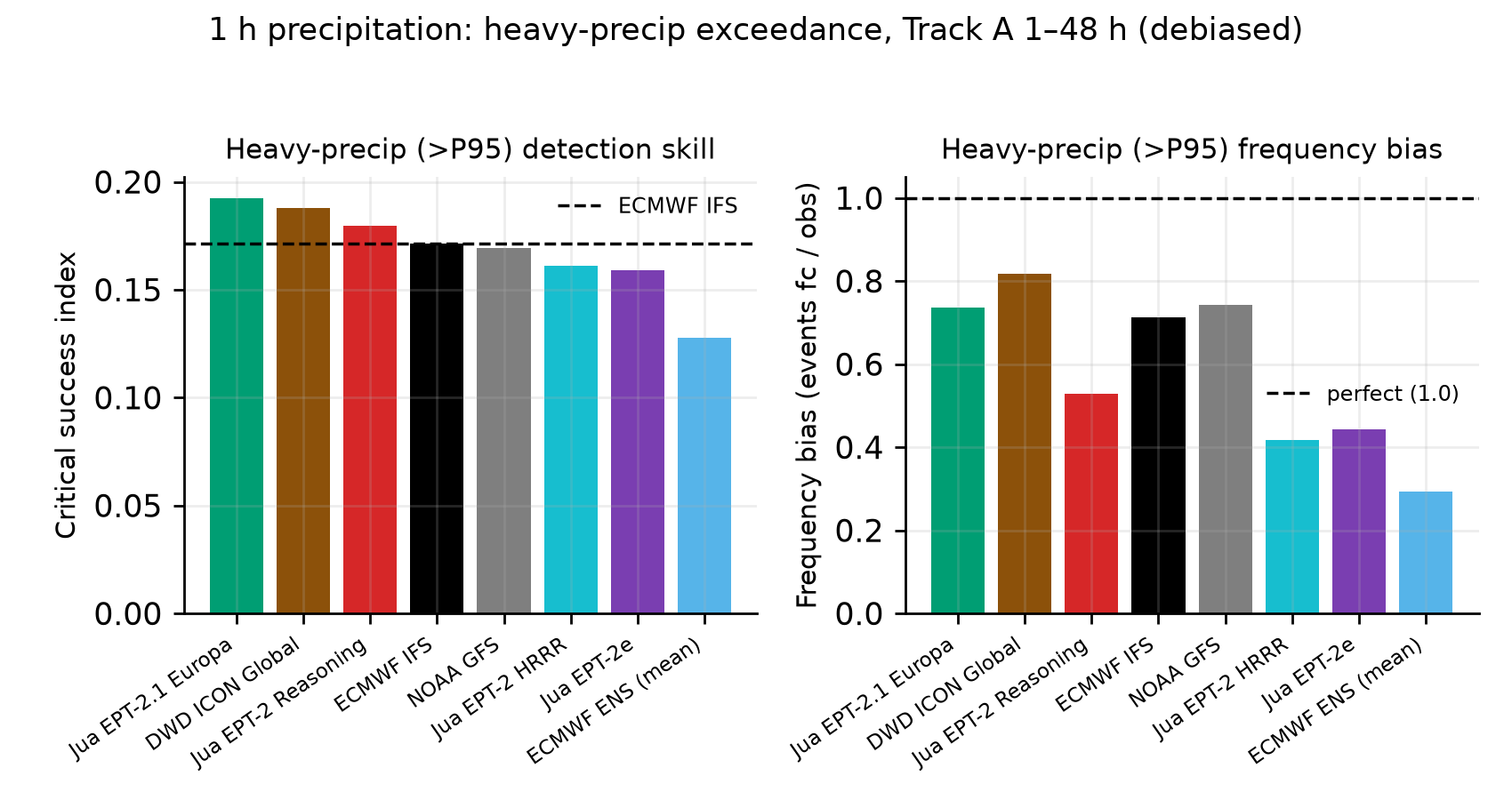}
\caption{Detection of heavy precipitation
(\(y\geq\mathrm{P95}\)) on Track~A over 1--48\,h.
Left: critical success index, with ECMWF IFS shown by the dashed line.
Right: frequency bias, with unity denoting the correct event frequency.
All models underpredict the frequency of heavy precipitation.}
\label{fig:precipcat}
\end{figure}

Figures~\ref{fig:leadwind}, \ref{fig:leadtemp}, \ref{fig:leadsolar}, and
\ref{fig:preciplead} show skill as a function of forecast lead.
Figure~\ref{fig:bars12} repeats the regime comparison over 1--12\,h, while
Table~\ref{tab:countryheat} reports heat-regime skill by country.
Figure~\ref{fig:precipbars12}, Table~\ref{tab:precipskill}, and
Figure~\ref{fig:precipcat} provide the short-range, full-horizon, and
categorical precipitation checks; short-range precipitation values are in
Table~\ref{tab:precipskillshort}.

\begin{table}[htbp]
\centering
\caption{Track~A MAE skill (\%) by country in the heat regime
(\(y\geq\mathrm{P95}\)), aggregated over 6--48\,h after bias correction.
NOAA GFS is negative in every country with available observations.}
\label{tab:countryheat}
\numtable{\resizebox{\textwidth}{!}{
\begin{tabular}{lrrrrrrrrrrrrr}
\toprule
Model & DE & FR & GB & ES & IT & PL & NL & BE & CH & NO & AT & CZ & DK \\
\midrule
Jua EPT-2.1 Europa & +6.6 & +16.1 & -1.2 & +6.4 & +17.2 & +6.2 & -3.4 & -3.0 & +31.1 & +5.5 & +28.2 & +16.6 & -4.4 \\
Jua EPT-2 HRRR & +10.5 & +21.5 & +4.0 & +10.5 & +27.6 & +5.6 & +1.8 & -0.6 & +29.4 & +14.2 & +16.6 & +17.7 & +12.2 \\
Jua EPT-2 Reasoning & +6.6 & +2.6 & +4.3 & -2.6 & +0.1 & +3.9 & +3.8 & -1.0 & -1.3 & -5.2 & -3.1 & +7.5 & +2.7 \\
Jua EPT-2e & +6.1 & -2.2 & -0.3 & -15.1 & -10.4 & -2.4 & +0.6 & -2.8 & -2.5 & -12.2 & -16.3 & -1.3 & +1.7 \\
Microsoft Aurora & +2.1 & -4.5 & -7.5 & -18.7 & -7.8 & -0.9 & +0.7 & -11.8 & -3.3 & -18.0 & -15.4 & -7.0 & +1.8 \\
ECMWF AIFS & +0.8 & -1.6 & -10.2 & -12.7 & -14.6 & +1.9 & -0.4 & +6.3 & +1.9 & -3.4 & -25.0 & -3.1 & +10.3 \\
ECMWF AIFS ENS (mean) & +5.3 & -3.1 & -6.7 & -10.6 & -12.6 & -3.1 & -3.9 & +5.1 & +4.6 & -3.2 & -38.7 & +2.1 & +4.4 \\
ECMWF ENS (mean) & -2.1 & -4.6 & -5.1 & -12.1 & -4.0 & -8.6 & +3.4 & -2.9 & -10.7 & -9.8 & -10.6 & -1.5 & -9.2 \\
NOAA GFS & -18.4 & -23.6 & -37.0 & -25.1 & -13.2 & -24.3 & -21.9 & -12.2 & -10.2 & -43.1 & -25.3 & -31.4 & -24.7 \\
DWD ICON Global & +6.9 & +8.8 & -4.5 & +3.6 & +12.8 & +4.8 & -2.7 & +10.0 & +22.7 & +18.1 & +16.8 & +15.5 & -6.5 \\
DWD ICON-EU$^{\dagger}$ & +13.1 & +18.4 & +1.3 & +7.4 & +20.1 & +9.1 & -2.6 & +14.1 & +34.6 & +6.6 & +28.1 & +17.8 & -0.7 \\
\bottomrule
\end{tabular}
}}
\end{table}

\begin{table}[htbp]
\centering
\caption{Track~A MAE skill (\%) relative to ECMWF IFS for hourly
precipitation by observed-value regime, aggregated over 1--48\,h after
multiplicative bias correction. Parentheses give the leave-one-month-out
jackknife SE.}
\label{tab:precipskill}
\numtable{\resizebox{\textwidth}{!}{
\begin{tabular}{lrrrrrr}
\toprule
Model & All & Dry & $<$P50 & P50--75 & P75--95 & $>$P95 \\
\midrule
Jua EPT-2 Reasoning & +5.1 ($\pm$0.4) & +2.2 ($\pm$1.3) & +9.3 ($\pm$0.6) & +14.0 ($\pm$0.9) & +10.6 ($\pm$0.8) & +1.7 ($\pm$0.5) \\
Jua EPT-2 HRRR & +1.3 ($\pm$1.3) & -6.1 ($\pm$3.3) & +9.3 ($\pm$1.8) & +15.2 ($\pm$2.2) & +10.2 ($\pm$1.2) & -1.5 ($\pm$0.5) \\
Jua EPT-2e & +3.3 ($\pm$1.1) & +1.5 ($\pm$4.8) & +7.7 ($\pm$1.1) & +14.8 ($\pm$1.7) & +8.8 ($\pm$1.5) & -1.9 ($\pm$0.9) \\
Jua EPT-2.1 Europa & +4.2 ($\pm$0.7) & +21.0 ($\pm$2.7) & -11.1 ($\pm$3.0) & -10.5 ($\pm$3.4) & -5.9 ($\pm$2.5) & +0.2 ($\pm$0.6) \\
NOAA GFS & -4.3 ($\pm$1.0) & -0.2 ($\pm$1.8) & -5.1 ($\pm$2.0) & -9.3 ($\pm$1.8) & -11.2 ($\pm$1.4) & -3.7 ($\pm$1.2) \\
DWD ICON Global & -1.4 ($\pm$1.7) & +8.3 ($\pm$2.7) & -14.4 ($\pm$2.7) & -14.4 ($\pm$3.0) & -9.8 ($\pm$2.2) & +0.0 ($\pm$0.4) \\
\bottomrule
\end{tabular}
}}
\end{table}

\FloatBarrier
\subsection{Matched regional comparison}
\label{app:trackb}

\begin{figure}[htbp]
\centering
\includegraphics[width=0.9\textwidth]{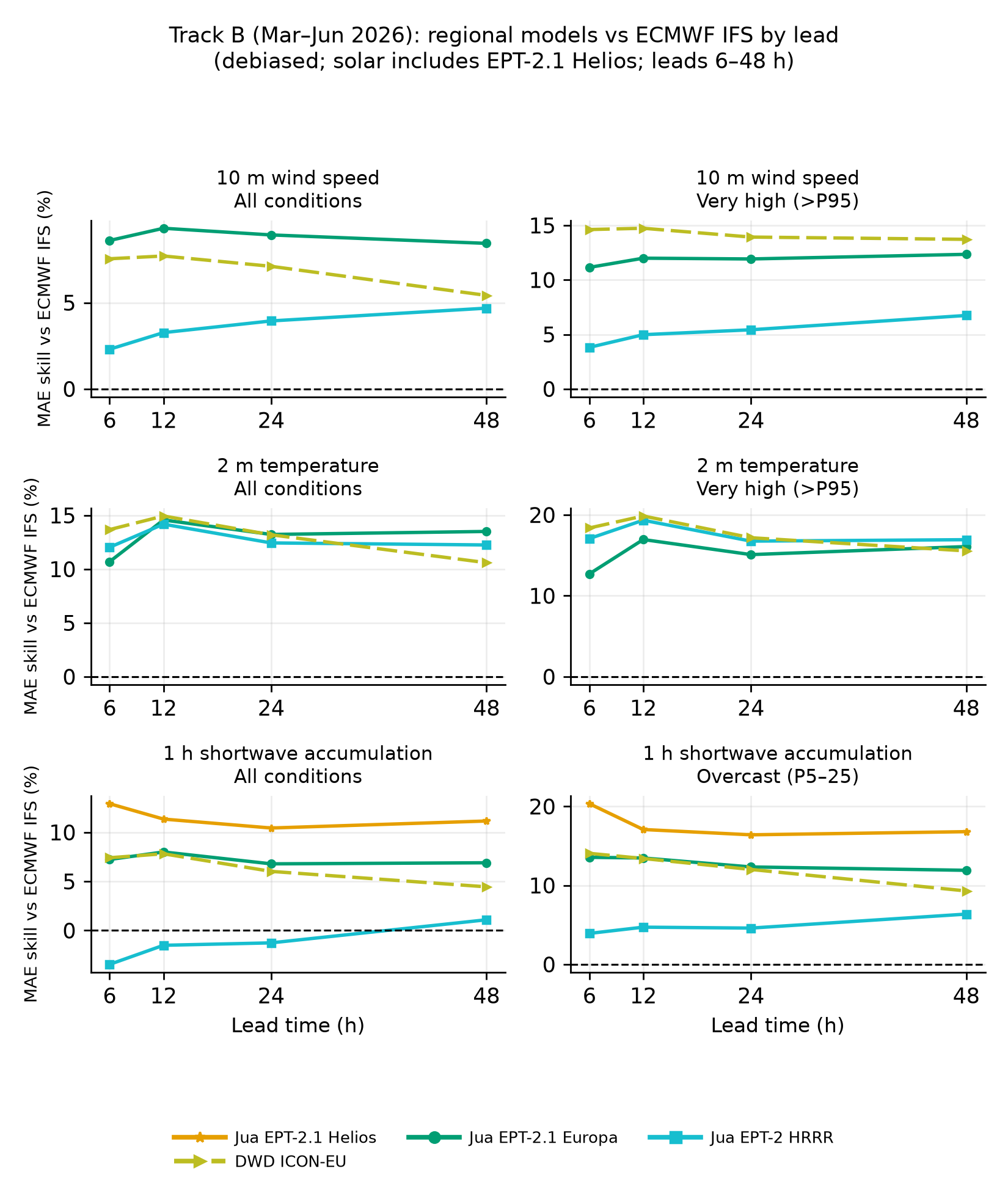}
\caption{Track~B MAE skill relative to ECMWF IFS by lead time. Rows show
10\,m wind, 2\,m temperature, and hourly solar radiation. Columns compare
all conditions with gale-force wind, heat, and overcast conditions,
respectively. Solar radiation additionally includes EPT-2.1 Helios.
Plotted leads are 6, 12, 24, and 48\,h; shading denotes
\(\pm1\) jackknife SE.}
\label{fig:trackb}
\end{figure}

\begin{figure}[htbp]
\centering
\includegraphics[width=\textwidth]{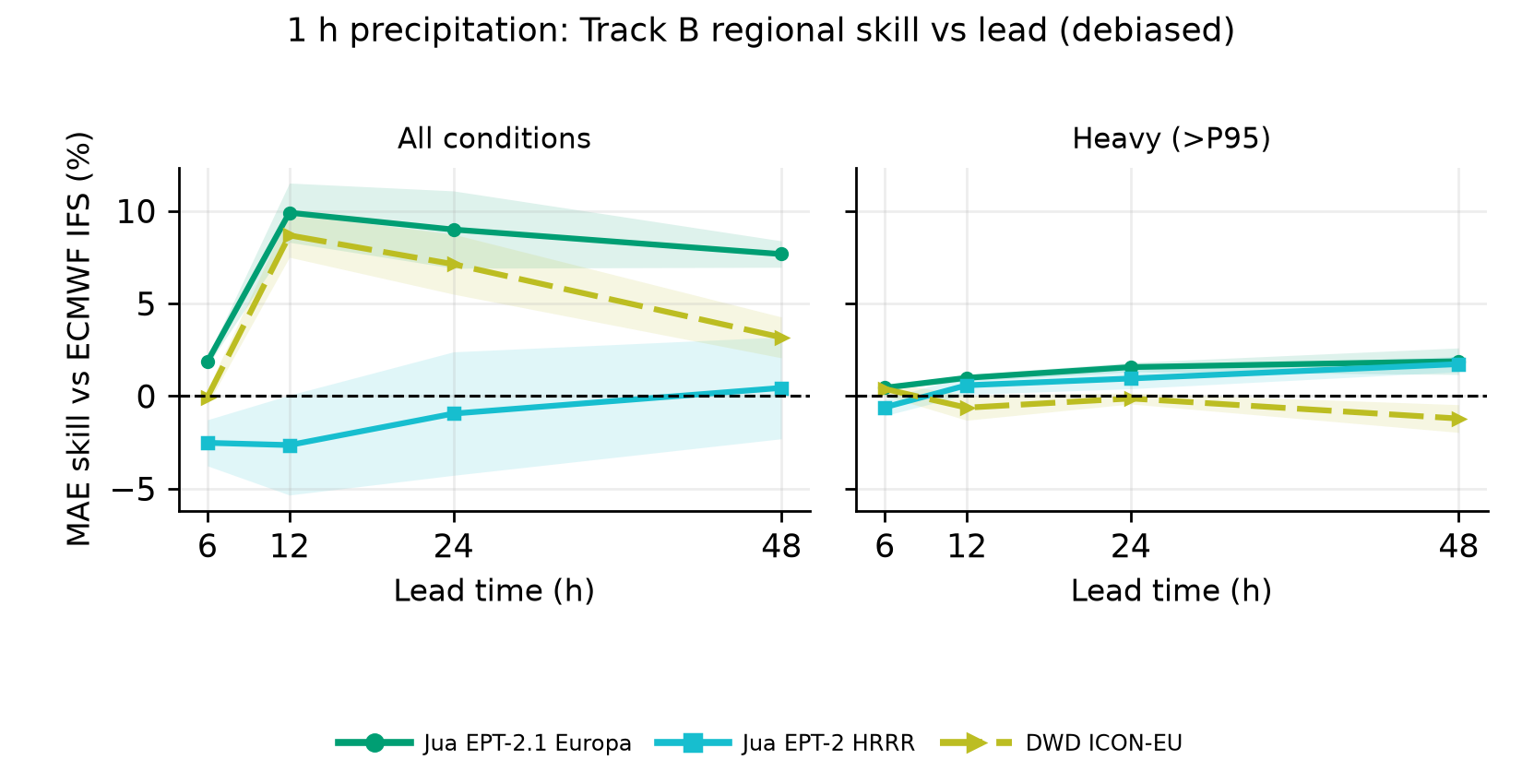}
\caption{Track~B precipitation skill relative to ECMWF IFS by lead time.
The left panel shows all conditions; the right panel shows the heavy
regime, \(y\geq\mathrm{P95}\). Shading denotes \(\pm1\)
jackknife SE.}
\label{fig:preciptrackblead}
\end{figure}

\begin{figure}[htbp]
\centering
\includegraphics[width=0.75\textwidth]{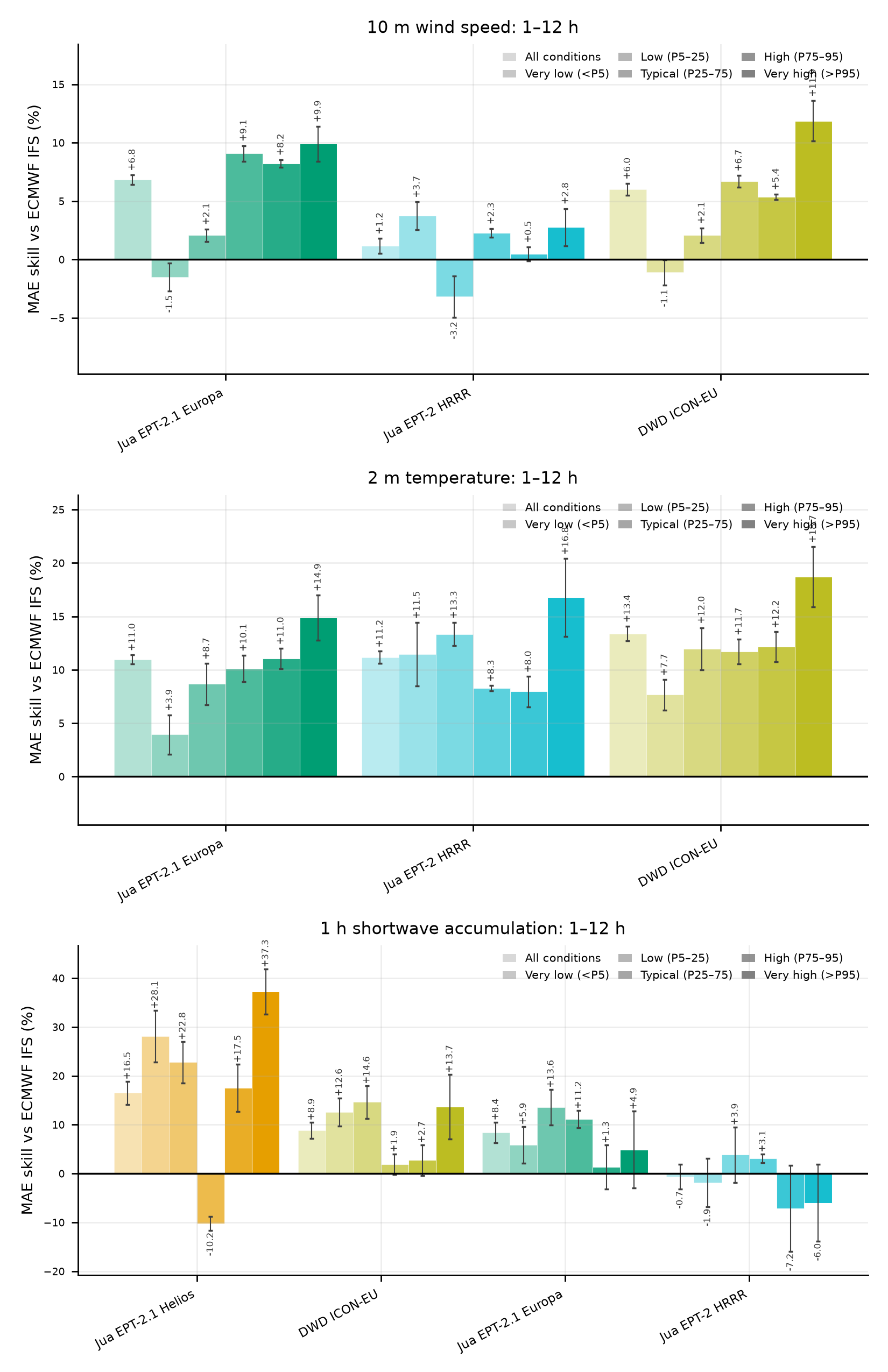}
\caption{Track~B MAE skill relative to ECMWF IFS for wind, temperature and solar by observed-value regime
over 1--12\,h, with \(\pm1\) jackknife SE}
\label{fig:trackbbars12}
\end{figure}

\begin{figure}[htbp]
\centering
\includegraphics[width=\textwidth]{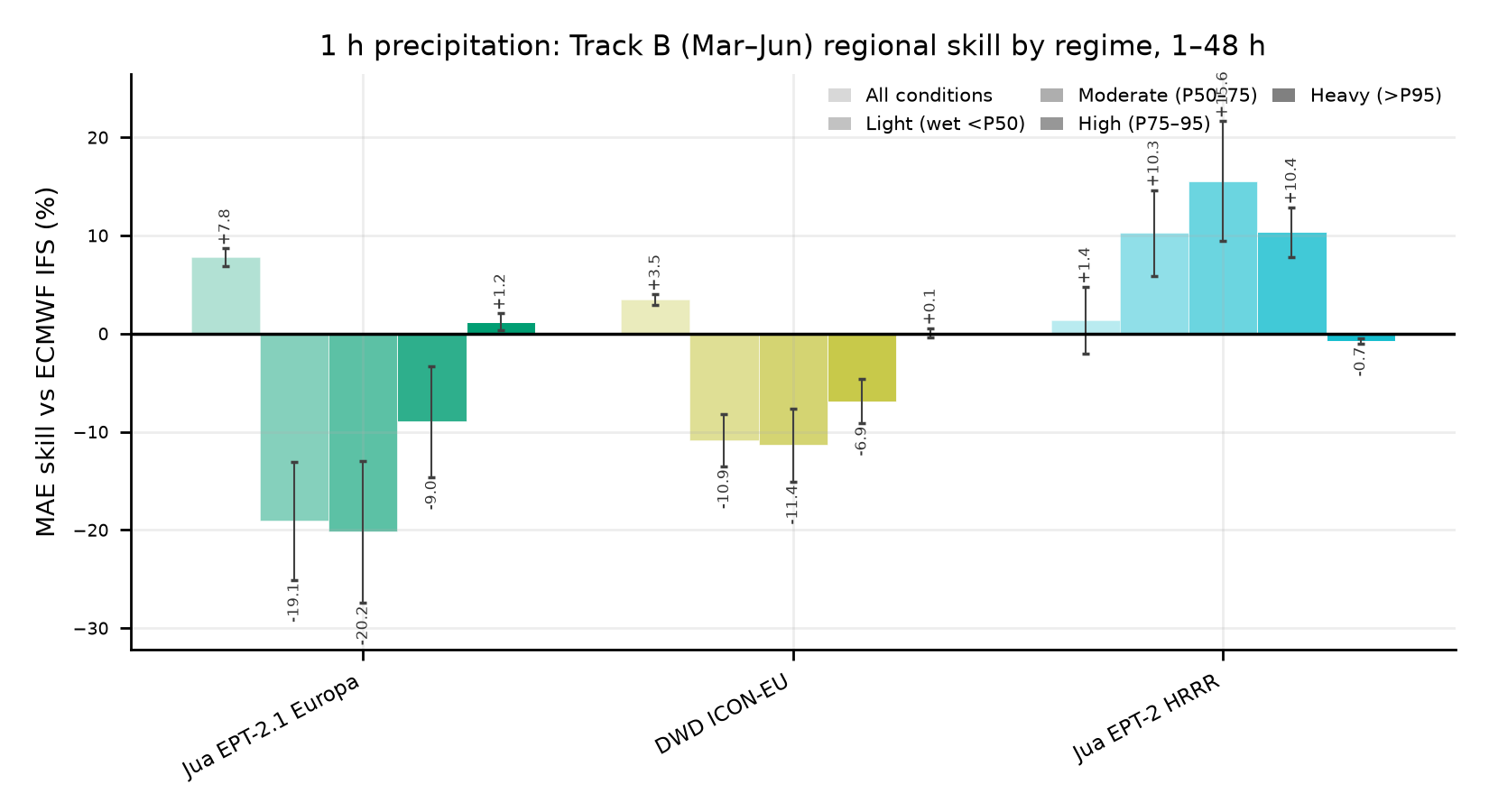}
\caption{Track~B MAE skill relative to ECMWF IFS for hourly precipitation
by observed-value regime, aggregated over 1--48\,h from March to June
2026, with \(\pm1\) jackknife SE.}
\label{fig:preciptrackb}
\end{figure}

Track~B robustness beyond the full-horizon comparison in the main text
is shown in Figure~\ref{fig:trackbbars12} for 1--12\,h and in
Figure~\ref{fig:trackb} as a function of lead.
Figures~\ref{fig:preciptrackb} and \ref{fig:preciptrackblead} give the
corresponding precipitation results.

\FloatBarrier
\subsection{Complete numerical results}
\label{app:numerical}

Track~A full-horizon regime values are in
Tables~\ref{tab:skillwind}, \ref{tab:skilltemp}, and
\ref{tab:skillsolar}; 1--12\,h values are in
Tables~\ref{tab:skillshort}, \ref{tab:skillshorttemp},
\ref{tab:skillshortsolar}, and \ref{tab:precipskillshort}; per-lead values
are in Tables~\ref{tab:leadwind}, \ref{tab:leadtemp}, and
\ref{tab:leadsolar}. Track~B regime values are in
Tables~\ref{tab:trackb}, \ref{tab:trackbtemp}, \ref{tab:trackbsolar}, and
\ref{tab:preciptrackb}; its per-lead values are in
Tables~\ref{tab:trackbleadwind}, \ref{tab:trackbleadtemp},
\ref{tab:trackbleadsolar}, and \ref{tab:preciptrackblead}.
Table~\ref{tab:precipcat} gives the heavy-precipitation detection metrics.

\begin{table}[htbp]
\centering
\caption{Track~A MAE skill (\%), 6--48\,h, by regime -- Figure~\ref{fig:bars},
wind. Parentheses: jackknife SE.}
\label{tab:skillwind}
\numtable{\resizebox{\textwidth}{!}{\begin{tabular}{@{}lrrrrrr@{}}
\toprule
Model & All & $<$P5 & P5--25 & P25--75 & P75--95 & $>$P95 \\
\midrule
Jua EPT-2.1 Europa & +8.4\,(0.5) & -1.7\,(1.4) & +6.4\,(0.9) & +11.3\,(0.7) & +8.5\,(0.4) & +9.0\,(1.2) \\\addlinespace[0.35em]
Jua EPT-2 HRRR & +4.5\,(0.3) & +2.7\,(1.0) & +1.4\,(0.9) & +5.1\,(0.3) & +4.9\,(1.0) & +6.1\,(0.9) \\\addlinespace[0.35em]
Jua EPT-2 Reasoning & +3.0\,(0.1) & -0.7\,(0.4) & +2.8\,(0.5) & +5.6\,(0.3) & +4.5\,(0.3) & -0.7\,(0.3) \\\addlinespace[0.35em]
Jua EPT-2e & -0.0\,(0.2) & -9.1\,(1.0) & -1.4\,(1.1) & +5.6\,(0.5) & +2.9\,(0.6) & -6.6\,(0.5) \\\addlinespace[0.35em]
Microsoft Aurora & -0.7\,(0.3) & -6.6\,(1.1) & -3.9\,(1.1) & +1.9\,(0.8) & +2.6\,(0.8) & -3.4\,(0.6) \\\addlinespace[0.35em]
ECMWF AIFS & -5.4\,(0.4) & -9.8\,(1.2) & -8.0\,(1.3) & -2.1\,(0.6) & -1.4\,(0.8) & -10.6\,(0.6) \\\addlinespace[0.35em]
ECMWF AIFS ENS (mean) & -2.3\,(0.5) & -12.0\,(1.2) & -3.3\,(1.5) & +3.8\,(0.7) & -0.1\,(0.7) & -9.0\,(0.7) \\\addlinespace[0.35em]
ECMWF ENS (mean) & +1.4\,(0.2) & -0.9\,(0.7) & +1.5\,(0.6) & +3.1\,(0.4) & +2.1\,(0.4) & -1.0\,(0.2) \\\addlinespace[0.35em]
NOAA GFS & -11.7\,(0.7) & +1.8\,(1.4) & -9.5\,(1.4) & -21.8\,(1.4) & -16.3\,(1.3) & +0.6\,(0.7) \\\addlinespace[0.35em]
DWD ICON Global & +3.6\,(0.4) & -4.1\,(1.2) & +2.8\,(1.0) & +2.6\,(0.7) & +2.5\,(0.7) & +9.5\,(1.1) \\
\bottomrule
\end{tabular}
}}
\end{table}

\begin{table}[htbp]
\centering
\caption{Track~A MAE skill (\%), 6--48\,h -- Figure~\ref{fig:bars}, temperature.}
\label{tab:skilltemp}
\numtable{\resizebox{\textwidth}{!}{\begin{tabular}{@{}lrrrrrr@{}}
\toprule
Model & All & $<$P5 & P5--25 & P25--75 & P75--95 & $>$P95 \\
\midrule
Jua EPT-2.1 Europa & +11.5\,(0.7) & +4.7\,(1.0) & +10.8\,(1.3) & +11.0\,(1.0) & +11.8\,(1.1) & +15.0\,(1.7) \\\addlinespace[0.35em]
Jua EPT-2 HRRR & +12.1\,(0.4) & +9.5\,(2.4) & +11.5\,(1.2) & +9.4\,(0.9) & +11.4\,(1.0) & +19.6\,(2.2) \\\addlinespace[0.35em]
Jua EPT-2 Reasoning & +4.6\,(0.4) & -6.9\,(1.3) & +2.6\,(1.3) & +7.4\,(0.4) & +7.0\,(0.8) & +1.1\,(0.7) \\\addlinespace[0.35em]
Jua EPT-2e & +0.8\,(0.7) & -24.3\,(3.6) & -5.0\,(2.8) & +6.1\,(0.8) & +7.4\,(1.1) & -5.0\,(0.7) \\\addlinespace[0.35em]
Microsoft Aurora & -1.3\,(1.1) & -22.2\,(4.4) & -7.0\,(4.6) & +4.1\,(1.0) & +3.8\,(1.6) & -6.9\,(1.0) \\\addlinespace[0.35em]
ECMWF AIFS & -1.4\,(0.6) & -27.3\,(3.6) & -5.1\,(2.3) & +2.3\,(1.0) & +4.9\,(1.4) & -4.9\,(2.0) \\\addlinespace[0.35em]
ECMWF AIFS ENS (mean) & +0.2\,(0.6) & -24.9\,(3.2) & -3.1\,(2.0) & +4.5\,(0.9) & +6.2\,(0.9) & -5.1\,(1.7) \\\addlinespace[0.35em]
ECMWF ENS (mean) & -1.7\,(0.3) & -12.2\,(1.0) & -3.3\,(0.9) & +1.4\,(0.4) & +0.3\,(0.8) & -6.1\,(0.5) \\\addlinespace[0.35em]
NOAA GFS & -23.9\,(1.4) & -25.3\,(3.5) & -26.3\,(2.9) & -26.5\,(2.5) & -17.8\,(2.1) & -22.8\,(2.0) \\\addlinespace[0.35em]
DWD ICON Global & +5.5\,(0.9) & -3.8\,(1.2) & +3.0\,(1.7) & +4.1\,(1.5) & +8.3\,(0.9) & +10.2\,(2.0) \\
\bottomrule
\end{tabular}
}}
\end{table}

\begin{table}[htbp]
\centering
\caption{Track~A MAE skill (\%), 1--48\,h hourly -- Figure~\ref{fig:bars},
solar. Parentheses: jackknife SE.}
\label{tab:skillsolar}
\numtable{\begin{tabular}{@{}lrrrrrr@{}}
\toprule
Model & All & $<$P5 & P5--25 & P25--75 & P75--95 & $>$P95 \\
\midrule
Jua EPT-2.1 Helios & +10.2\,(1.7) & +15.4\,(5.0) & +16.4\,(3.4) & -8.2\,(1.1) & +13.7\,(2.9) & +24.8\,(5.4) \\\addlinespace[0.35em]
Jua EPT-2.1 Europa & +7.1\,(1.0) & +3.1\,(4.0) & +11.0\,(3.1) & +9.2\,(2.3) & +1.9\,(4.3) & +4.4\,(4.8) \\\addlinespace[0.35em]
Jua EPT-2 HRRR & +0.6\,(1.4) & -3.6\,(4.6) & +3.8\,(3.7) & +4.5\,(1.0) & -4.6\,(4.8) & -4.1\,(4.9) \\\addlinespace[0.35em]
Jua EPT-2 Reasoning & +6.3\,(0.5) & +0.9\,(1.3) & +9.5\,(0.8) & +14.0\,(0.7) & -0.8\,(1.9) & -2.6\,(2.6) \\\addlinespace[0.35em]
Jua EPT-2e & +2.0\,(1.4) & -10.6\,(3.3) & +5.9\,(2.3) & +13.5\,(1.6) & -7.5\,(4.1) & -7.6\,(6.0) \\\addlinespace[0.35em]
NOAA GFS & -11.2\,(1.6) & -3.3\,(4.4) & -24.3\,(2.1) & -28.3\,(2.5) & +8.8\,(3.4) & +16.7\,(4.2) \\\addlinespace[0.35em]
DWD ICON Global & +5.6\,(0.8) & +6.5\,(3.3) & +8.1\,(2.6) & +3.1\,(1.5) & +4.2\,(3.3) & +6.9\,(4.2) \\
\bottomrule
\end{tabular}
}
\end{table}

\begin{table}[htbp]
\centering
\caption{Track~A MAE skill (\%), 1--12\,h -- Figure~\ref{fig:bars12}, wind.}
\label{tab:skillshort}
\numtable{\begin{tabular}{@{}lrrrrrr@{}}
\toprule
Model & All & $<$P5 & P5--25 & P25--75 & P75--95 & $>$P95 \\
\midrule
Jua EPT-2.1 Europa & +7.0\,(0.5) & -3.7\,(1.0) & +3.6\,(1.1) & +10.5\,(0.7) & +8.1\,(0.3) & +8.2\,(0.9) \\\addlinespace[0.35em]
Jua EPT-2 HRRR & +2.0\,(0.3) & +0.7\,(1.0) & -2.1\,(0.8) & +2.4\,(0.2) & +3.0\,(0.9) & +4.7\,(0.9) \\\addlinespace[0.35em]
Jua EPT-2 Reasoning & +1.5\,(0.1) & -1.5\,(0.2) & +0.4\,(0.3) & +3.2\,(0.2) & +3.4\,(0.2) & -0.4\,(0.2) \\\addlinespace[0.35em]
Jua EPT-2e & -2.4\,(0.2) & -10.8\,(0.7) & -6.0\,(0.7) & +2.1\,(0.5) & +1.7\,(0.4) & -6.2\,(0.4) \\\addlinespace[0.35em]
ECMWF ENS (mean) & +0.4\,(0.3) & -- & -- & -- & -- & -- \\\addlinespace[0.35em]
NOAA GFS & -10.9\,(0.8) & +4.8\,(0.8) & -5.6\,(0.9) & -21.5\,(1.8) & -18.1\,(1.4) & +0.7\,(0.6) \\\addlinespace[0.35em]
DWD ICON Global & +3.7\,(0.3) & -1.3\,(0.6) & +4.0\,(0.9) & +3.3\,(0.7) & +1.5\,(0.4) & +8.1\,(1.1) \\
\bottomrule
\end{tabular}
}
\end{table}

\begin{table}[htbp]
\centering
\caption{Track~A MAE skill (\%), 1--12\,h -- Figure~\ref{fig:bars12}, temperature.}
\label{tab:skillshorttemp}
\numtable{\begin{tabular}{@{}lrrrrrr@{}}
\toprule
Model & All & $<$P5 & P5--25 & P25--75 & P75--95 & $>$P95 \\
\midrule
Jua EPT-2.1 Europa & +8.3\,(0.8) & +2.9\,(1.1) & +7.3\,(1.0) & +7.8\,(1.1) & +7.5\,(1.1) & +13.3\,(1.4) \\\addlinespace[0.35em]
Jua EPT-2 HRRR & +9.5\,(0.6) & +9.0\,(2.3) & +8.7\,(1.6) & +6.2\,(1.0) & +8.2\,(0.8) & +18.8\,(2.0) \\\addlinespace[0.35em]
Jua EPT-2 Reasoning & +2.4\,(0.2) & -5.3\,(0.6) & +1.0\,(0.6) & +4.3\,(0.2) & +4.2\,(0.5) & +0.0\,(0.5) \\\addlinespace[0.35em]
Jua EPT-2e & -2.5\,(0.6) & -24.3\,(2.7) & -7.7\,(2.2) & +2.3\,(0.7) & +3.0\,(1.1) & -7.2\,(0.5) \\\addlinespace[0.35em]
ECMWF ENS (mean) & -3.0\,(0.2) & -- & -- & -- & -- & -- \\\addlinespace[0.35em]
NOAA GFS & -26.9\,(1.5) & -29.6\,(4.0) & -32.4\,(3.5) & -29.0\,(2.4) & -20.5\,(1.6) & -23.9\,(2.5) \\\addlinespace[0.35em]
DWD ICON Global & +5.3\,(0.8) & -2.9\,(1.0) & +2.8\,(1.2) & +4.9\,(1.5) & +6.3\,(0.7) & +10.1\,(1.4) \\
\bottomrule
\end{tabular}
}
\end{table}

\begin{table}[htbp]
\centering
\caption{Track~A MAE skill (\%), 1--12\,h -- Figure~\ref{fig:bars12}, solar.}
\label{tab:skillshortsolar}
\numtable{\begin{tabular}{@{}lrrrrrr@{}}
\toprule
Model & All & $<$P5 & P5--25 & P25--75 & P75--95 & $>$P95 \\
\midrule
Jua EPT-2.1 Helios & +13.5\,(1.9) & +20.6\,(5.9) & +20.1\,(2.9) & -6.9\,(1.5) & +17.7\,(2.6) & +28.7\,(6.1) \\\addlinespace[0.35em]
Jua EPT-2.1 Europa & +6.5\,(1.4) & +1.2\,(3.9) & +10.5\,(3.8) & +8.1\,(3.0) & +1.9\,(4.3) & +5.6\,(4.8) \\\addlinespace[0.35em]
Jua EPT-2 HRRR & -2.1\,(1.8) & -6.8\,(4.7) & +0.9\,(4.3) & +0.3\,(1.5) & -6.4\,(5.0) & -3.7\,(4.8) \\\addlinespace[0.35em]
Jua EPT-2 Reasoning & +4.4\,(0.4) & +0.6\,(1.3) & +7.4\,(0.8) & +11.0\,(0.7) & -2.4\,(1.3) & -3.0\,(1.7) \\\addlinespace[0.35em]
Jua EPT-2e & -1.1\,(1.7) & -13.5\,(3.4) & +4.1\,(3.1) & +11.0\,(1.7) & -13.8\,(4.5) & -10.0\,(5.0) \\\addlinespace[0.35em]
NOAA GFS & -12.7\,(1.9) & -5.8\,(4.2) & -27.6\,(2.7) & -29.8\,(2.6) & +9.7\,(4.2) & +15.9\,(3.6) \\\addlinespace[0.35em]
DWD ICON Global & +6.4\,(1.2) & +6.2\,(3.4) & +9.7\,(3.0) & +3.4\,(1.6) & +4.1\,(2.9) & +8.8\,(4.8) \\
\bottomrule
\end{tabular}
}
\end{table}

\begin{table}[htbp]
\centering
\caption{Track~A precipitation MAE skill (\%), 1--12\,h
(Figure~\ref{fig:precipbars12}). Parentheses: jackknife SE.}
\label{tab:precipskillshort}
\numtable{\resizebox{\textwidth}{!}{\begin{tabular}{lrrrrrr}
\toprule
Model & All & Dry & $<$P50 & P50--75 & P75--95 & $>$P95 \\
\midrule
Jua EPT-2 Reasoning & +1.9 ($\pm$0.5) & -2.9 ($\pm$1.0) & +3.3 ($\pm$1.0) & +8.2 ($\pm$1.0) & +7.8 ($\pm$0.7) & +2.2 ($\pm$0.5) \\
Jua EPT-2 HRRR & -2.7 ($\pm$1.1) & -10.9 ($\pm$3.8) & +3.9 ($\pm$1.2) & +10.8 ($\pm$1.5) & +6.8 ($\pm$1.1) & -4.5 ($\pm$0.8) \\
Jua EPT-2e & +4.6 ($\pm$0.9) & +12.4 ($\pm$3.4) & +8.9 ($\pm$1.1) & +11.2 ($\pm$1.1) & +2.2 ($\pm$1.5) & -6.0 ($\pm$1.3) \\
Jua EPT-2.1 Europa & +2.2 ($\pm$1.5) & +22.0 ($\pm$3.2) & -16.1 ($\pm$4.2) & -16.5 ($\pm$4.6) & -10.8 ($\pm$3.2) & -1.1 ($\pm$0.7) \\
NOAA GFS & -7.3 ($\pm$1.0) & -3.4 ($\pm$3.6) & -9.1 ($\pm$1.4) & -14.2 ($\pm$1.5) & -16.1 ($\pm$1.9) & -4.8 ($\pm$1.8) \\
DWD ICON Global & -1.4 ($\pm$2.3) & +11.4 ($\pm$3.3) & -17.4 ($\pm$3.8) & -17.1 ($\pm$4.5) & -11.5 ($\pm$3.5) & -0.3 ($\pm$0.6) \\
\bottomrule
\end{tabular}
}}
\end{table}

\begin{table}[htbp]
\centering
\caption{Track~A MAE skill (\%) by lead -- Figure~\ref{fig:leadwind}, wind.
Point estimates (no SE in per-lead tables).}
\label{tab:leadwind}
\numtable{\begin{tabular}{@{}l*{8}{r}@{}}
\toprule
& \multicolumn{2}{c}{6\,h} & \multicolumn{2}{c}{12\,h} & \multicolumn{2}{c}{24\,h} & \multicolumn{2}{c}{48\,h} \\
\cmidrule(lr){2-3} \cmidrule(lr){4-5} \cmidrule(lr){6-7} \cmidrule(lr){8-9}
Model & All & $>$P95 & All & $>$P95 & All & $>$P95 & All & $>$P95 \\
\midrule
EPT-2.1 Europa & +8.3 & +8.8 & +8.7 & +9.3 & +8.5 & +9.2 & +8.1 & +9.0 \\\addlinespace[0.3em]
EPT-2 HRRR & +3.3 & +5.0 & +3.9 & +5.6 & +4.6 & +6.0 & +5.1 & +6.9 \\\addlinespace[0.3em]
EPT-2 Reasoning & +1.9 & -0.7 & +2.3 & -0.8 & +2.9 & -0.9 & +3.9 & -0.4 \\\addlinespace[0.3em]
EPT-2e & -1.2 & -8.0 & -1.0 & -7.5 & -0.1 & -6.7 & +1.1 & -5.5 \\\addlinespace[0.3em]
Aurora & -1.6 & -4.0 & -1.6 & -3.6 & -0.9 & -3.6 & +0.4 & -2.8 \\\addlinespace[0.3em]
AIFS & -6.6 & -10.8 & -6.5 & -11.0 & -5.7 & -11.1 & -4.0 & -9.5 \\\addlinespace[0.3em]
AIFS ENS & -3.5 & -8.3 & -3.4 & -8.9 & -2.7 & -9.4 & -0.7 & -8.9 \\\addlinespace[0.3em]
ENS & +0.3 & -0.9 & +0.7 & -0.8 & +1.2 & -0.9 & +2.3 & -1.5 \\\addlinespace[0.3em]
GFS & -11.1 & +2.0 & -11.1 & +1.8 & -11.8 & +0.6 & -12.3 & -0.5 \\\addlinespace[0.3em]
ICON Global & +4.6 & +10.1 & +4.6 & +10.1 & +4.0 & +9.6 & +2.4 & +8.6 \\
\bottomrule
\end{tabular}
}
\end{table}

\begin{table}[htbp]
\centering
\caption{Track~A MAE skill (\%) by lead -- Figure~\ref{fig:leadtemp}, temperature.
Point estimates (no SE in per-lead tables).}
\label{tab:leadtemp}
\numtable{\begin{tabular}{@{}l*{8}{r}@{}}
\toprule
& \multicolumn{2}{c}{6\,h} & \multicolumn{2}{c}{12\,h} & \multicolumn{2}{c}{24\,h} & \multicolumn{2}{c}{48\,h} \\
\cmidrule(lr){2-3} \cmidrule(lr){4-5} \cmidrule(lr){6-7} \cmidrule(lr){8-9}
Model & All & $>$P95 & All & $>$P95 & All & $>$P95 & All & $>$P95 \\
\midrule
EPT-2.1 Europa & +8.4 & +12.5 & +12.0 & +15.8 & +11.5 & +14.5 & +11.6 & +14.4 \\\addlinespace[0.3em]
EPT-2 HRRR & +10.1 & +19.5 & +12.6 & +21.3 & +12.1 & +19.4 & +12.2 & +18.4 \\\addlinespace[0.3em]
EPT-2 Reasoning & +2.2 & -0.5 & +3.4 & +0.2 & +4.5 & +0.9 & +6.0 & +2.2 \\\addlinespace[0.3em]
EPT-2e & -4.2 & -10.0 & -1.5 & -7.5 & +1.1 & -4.7 & +3.2 & -2.7 \\\addlinespace[0.3em]
Aurora & -4.6 & -7.8 & -2.3 & -7.1 & -1.2 & -7.0 & +0.2 & -6.7 \\\addlinespace[0.3em]
AIFS & -4.5 & -7.5 & -2.5 & -5.2 & -1.7 & -5.0 & +0.4 & -3.7 \\\addlinespace[0.3em]
AIFS ENS & -4.3 & -10.1 & -1.3 & -6.6 & +0.1 & -5.0 & +2.2 & -3.3 \\\addlinespace[0.3em]
ENS & -3.0 & -6.7 & -2.6 & -6.3 & -1.9 & -6.0 & -0.6 & -5.6 \\\addlinespace[0.3em]
GFS & -27.3 & -24.6 & -24.2 & -22.5 & -24.2 & -23.2 & -22.3 & -22.6 \\\addlinespace[0.3em]
ICON Global & +5.1 & +10.1 & +6.9 & +11.7 & +6.0 & +10.2 & +4.0 & +8.8 \\
\bottomrule
\end{tabular}
}
\end{table}

\begin{table}[htbp]
\centering
\caption{Track~A MAE skill (\%) by lead -- Figure~\ref{fig:leadsolar}, solar
(Ovc.\ = overcast, P5--P25). Point estimates; SEs in regime tables.}
\label{tab:leadsolar}
\numtable{\begin{tabular}{@{}l*{10}{r}@{}}
\toprule
& \multicolumn{2}{c}{1\,h} & \multicolumn{2}{c}{6\,h} & \multicolumn{2}{c}{12\,h} & \multicolumn{2}{c}{24\,h} & \multicolumn{2}{c}{48\,h} \\
\cmidrule(lr){2-3} \cmidrule(lr){4-5} \cmidrule(lr){6-7} \cmidrule(lr){8-9} \cmidrule(lr){10-11}
Model & All & Ovc. & All & Ovc. & All & Ovc. & All & Ovc. & All & Ovc. \\
\midrule
EPT-2.1 Helios & +26.1 & -- & +10.9 & +19.0 & +9.0 & +15.2 & +8.1 & +15.4 & +9.2 & +15.5 \\\addlinespace[0.3em]
EPT-2.1 Europa & +6.2 & -- & +5.6 & +10.1 & +6.7 & +10.0 & +6.1 & +10.2 & +6.3 & +10.2 \\\addlinespace[0.3em]
ICON Global & +7.1 & -- & +5.6 & +9.8 & +6.2 & +8.9 & +5.2 & +8.1 & +4.3 & +6.3 \\\addlinespace[0.3em]
EPT-2 Reasoning & +3.5 & -- & +4.2 & +7.0 & +5.1 & +8.5 & +6.0 & +9.5 & +7.8 & +11.1 \\\addlinespace[0.3em]
EPT-2e & +0.8 & -- & -4.5 & +2.4 & -2.5 & +2.2 & +0.2 & +2.9 & +4.1 & +8.1 \\\addlinespace[0.3em]
EPT-2 HRRR & -4.6 & -- & -4.8 & -0.5 & -2.7 & +1.3 & -1.7 & +2.6 & +1.1 & +5.1 \\\addlinespace[0.3em]
GFS & -12.8 & -- & -13.7 & -30.2 & -12.4 & -29.4 & -11.7 & -25.8 & -11.1 & -24.3 \\
\bottomrule
\end{tabular}
}
\end{table}

\begin{table}[htbp]
\centering
\caption{Track~B wind MAE skill (\%), 1--48\,h, by regime
(Figure~\ref{fig:trackbbars}).}
\label{tab:trackb}
\numtable{\begin{tabular}{@{}lrrrrrr@{}}
\toprule
Model & All & $<$P5 & P5--25 & P25--75 & P75--95 & $>$P95 \\
\midrule
Jua EPT-2.1 Europa & +7.2\,(0.4) & +0.1\,(0.8) & +3.0\,(0.3) & +8.9\,(0.6) & +8.2\,(0.3) & +10.5\,(1.4) \\\addlinespace[0.35em]
Jua EPT-2 HRRR & +2.4\,(0.6) & +4.9\,(1.0) & -1.8\,(1.7) & +3.5\,(0.4) & +1.8\,(0.4) & +4.1\,(1.5) \\\addlinespace[0.35em]
DWD ICON-EU & +5.3\,(0.5) & -0.7\,(1.0) & +1.8\,(0.6) & +5.6\,(0.5) & +3.9\,(0.2) & +11.5\,(1.6) \\
\bottomrule
\end{tabular}
}
\end{table}

\begin{table}[htbp]
\centering
\caption{Track~B MAE skill (\%), 1--48\,h -- Figure~\ref{fig:trackbbars},
temperature.}
\label{tab:trackbtemp}
\numtable{\begin{tabular}{@{}lrrrrrr@{}}
\toprule
Model & All & $<$P5 & P5--25 & P25--75 & P75--95 & $>$P95 \\
\midrule
Jua EPT-2.1 Europa & +13.2\,(0.6) & +5.9\,(1.8) & +11.7\,(2.0) & +12.3\,(1.3) & +13.5\,(0.6) & +16.6\,(2.3) \\\addlinespace[0.35em]
Jua EPT-2 HRRR & +12.0\,(0.7) & +12.8\,(3.3) & +14.9\,(1.4) & +9.2\,(0.4) & +8.7\,(1.3) & +17.2\,(3.5) \\\addlinespace[0.35em]
DWD ICON-EU & +12.4\,(0.8) & +7.7\,(1.1) & +11.8\,(2.3) & +10.6\,(1.1) & +11.3\,(1.5) & +17.2\,(3.2) \\
\bottomrule
\end{tabular}
}
\end{table}

\begin{table}[htbp]
\centering
\caption{Track~B MAE skill (\%), 1--48\,h -- Figure~\ref{fig:trackbbars},
solar.}
\label{tab:trackbsolar}
\numtable{\begin{tabular}{@{}lrrrrrr@{}}
\toprule
Model & All & $<$P5 & P5--25 & P25--75 & P75--95 & $>$P95 \\
\midrule
Jua EPT-2.1 Helios & +12.6\,(2.5) & +21.6\,(4.6) & +18.1\,(5.5) & -11.3\,(0.5) & +13.5\,(5.8) & +32.2\,(4.9) \\\addlinespace[0.35em]
Jua EPT-2.1 Europa & +7.8\,(2.0) & +7.4\,(3.9) & +12.8\,(3.1) & +10.7\,(1.7) & +0.1\,(4.5) & +3.4\,(7.8) \\\addlinespace[0.35em]
Jua EPT-2 HRRR & +0.8\,(2.2) & -0.2\,(5.1) & +4.8\,(5.7) & +6.5\,(0.3) & -6.1\,(8.2) & -6.4\,(7.7) \\\addlinespace[0.35em]
DWD ICON-EU & +7.2\,(1.4) & +12.4\,(3.3) & +12.4\,(2.5) & +1.0\,(2.2) & +0.9\,(3.3) & +10.5\,(5.7) \\
\bottomrule
\end{tabular}
}
\end{table}

\begin{table}[htbp]
\centering
\caption{Track~B MAE skill (\%) by lead -- Figure~\ref{fig:trackb}, wind.}
\label{tab:trackbleadwind}
\numtable{\begin{tabular}{@{}l*{8}{r}@{}}
\toprule
& \multicolumn{2}{c}{6\,h} & \multicolumn{2}{c}{12\,h} & \multicolumn{2}{c}{24\,h} & \multicolumn{2}{c}{48\,h} \\
\cmidrule(lr){2-3} \cmidrule(lr){4-5} \cmidrule(lr){6-7} \cmidrule(lr){8-9}
Model & All & $>$P95 & All & $>$P95 & All & $>$P95 & All & $>$P95 \\
\midrule
EPT-2.1 Europa & +8.6 & +11.1 & +9.3 & +12.0 & +8.9 & +11.9 & +8.4 & +12.3 \\\addlinespace[0.3em]
EPT-2 HRRR & +2.3 & +3.8 & +3.3 & +5.0 & +3.9 & +5.4 & +4.7 & +6.7 \\\addlinespace[0.3em]
ICON-EU & +7.5 & +14.6 & +7.7 & +14.7 & +7.1 & +13.9 & +5.4 & +13.7 \\
\bottomrule
\end{tabular}
}
\end{table}

\begin{table}[htbp]
\centering
\caption{Track~B MAE skill (\%) by lead -- Figure~\ref{fig:trackb},
temperature.}
\label{tab:trackbleadtemp}
\numtable{\begin{tabular}{@{}l*{8}{r}@{}}
\toprule
& \multicolumn{2}{c}{6\,h} & \multicolumn{2}{c}{12\,h} & \multicolumn{2}{c}{24\,h} & \multicolumn{2}{c}{48\,h} \\
\cmidrule(lr){2-3} \cmidrule(lr){4-5} \cmidrule(lr){6-7} \cmidrule(lr){8-9}
Model & All & $>$P95 & All & $>$P95 & All & $>$P95 & All & $>$P95 \\
\midrule
EPT-2.1 Europa & +10.7 & +12.7 & +14.6 & +17.0 & +13.2 & +15.1 & +13.5 & +16.1 \\\addlinespace[0.3em]
EPT-2 HRRR & +12.0 & +17.1 & +14.2 & +19.3 & +12.4 & +16.8 & +12.2 & +16.9 \\\addlinespace[0.3em]
ICON-EU & +13.7 & +18.4 & +14.9 & +19.9 & +13.2 & +17.2 & +10.6 & +15.5 \\
\bottomrule
\end{tabular}
}
\end{table}

\begin{table}[htbp]
\centering
\caption{Track~B MAE skill (\%) by lead -- Figure~\ref{fig:trackb}, solar
(Ovc.\ = overcast, P5--P25).}
\label{tab:trackbleadsolar}
\numtable{\begin{tabular}{@{}l*{8}{r}@{}}
\toprule
& \multicolumn{2}{c}{6\,h} & \multicolumn{2}{c}{12\,h} & \multicolumn{2}{c}{24\,h} & \multicolumn{2}{c}{48\,h} \\
\cmidrule(lr){2-3} \cmidrule(lr){4-5} \cmidrule(lr){6-7} \cmidrule(lr){8-9}
Model & All & Ovc. & All & Ovc. & All & Ovc. & All & Ovc. \\
\midrule
EPT-2.1 Helios & +12.9 & +20.3 & +11.4 & +17.1 & +10.4 & +16.4 & +11.2 & +16.8 \\\addlinespace[0.3em]
EPT-2.1 Europa & +7.2 & +13.6 & +8.0 & +13.5 & +6.8 & +12.3 & +6.9 & +11.9 \\\addlinespace[0.3em]
EPT-2 HRRR & -3.5 & +3.9 & -1.5 & +4.7 & -1.3 & +4.6 & +1.1 & +6.4 \\\addlinespace[0.3em]
ICON-EU & +7.4 & +14.1 & +7.8 & +13.4 & +6.0 & +12.0 & +4.4 & +9.3 \\
\bottomrule
\end{tabular}
}
\end{table}

\begin{table}[htbp]
\centering
\caption{Track~B precipitation MAE skill (\%) by lead
(Figure~\ref{fig:preciptrackblead}). Point estimates.}
\label{tab:preciptrackblead}
\numtable{\begin{tabular}{@{}l*{8}{r}@{}}
\toprule
& \multicolumn{2}{c}{6\,h} & \multicolumn{2}{c}{12\,h} & \multicolumn{2}{c}{24\,h} & \multicolumn{2}{c}{48\,h} \\
\cmidrule(lr){2-3} \cmidrule(lr){4-5} \cmidrule(lr){6-7} \cmidrule(lr){8-9}
Model & All & $>$P95 & All & $>$P95 & All & $>$P95 & All & $>$P95 \\
\midrule
Jua EPT-2.1 Europa & +1.9 & +0.5 & +9.9 & +1.0 & +9.0 & +1.6 & +7.7 & +1.9 \\
Jua EPT-2 HRRR & -2.5 & -0.6 & -2.6 & +0.6 & -0.9 & +1.0 & +0.4 & +1.7 \\
DWD ICON-EU & -0.1 & +0.4 & +8.7 & -0.6 & +7.1 & -0.1 & +3.2 & -1.2 \\
\bottomrule
\end{tabular}
}
\end{table}

\begin{table}[htbp]
\centering
\caption{Track~B MAE skill (\%) for 1\,h precipitation by wet-hour regime,
1 to 48\,h, debiased -- Figure~\ref{fig:preciptrackb}. Parentheses:
jackknife SE.}
\label{tab:preciptrackb}
\numtable{\resizebox{\textwidth}{!}{\begin{tabular}{lrrrrrr}
\toprule
Model & All & Dry & $<$P50 & P50--75 & P75--95 & $>$P95 \\
\midrule
Jua EPT-2.1 Europa & +7.8 ($\pm$0.9) & +29.3 ($\pm$1.4) & -19.1 ($\pm$6.0) & -20.2 ($\pm$7.2) & -9.0 ($\pm$5.7) & +1.2 ($\pm$0.8) \\
Jua EPT-2 HRRR & +1.4 ($\pm$3.4) & -3.6 ($\pm$9.7) & +10.3 ($\pm$4.4) & +15.6 ($\pm$6.1) & +10.4 ($\pm$2.5) & -0.7 ($\pm$0.3) \\
DWD ICON-EU & +3.5 ($\pm$0.5) & +15.3 ($\pm$0.3) & -10.9 ($\pm$2.7) & -11.4 ($\pm$3.8) & -6.9 ($\pm$2.2) & +0.1 ($\pm$0.5) \\
\bottomrule
\end{tabular}
}}
\end{table}
\begin{table}[htbp]
\centering
\caption{Heavy-precipitation ($>$P95) detection on Track~A, 1 to 48\,h,
debiased -- Figure~\ref{fig:precipcat}. POD: probability of detection; FAR:
false-alarm ratio; CSI: critical success index; frequency bias is forecast
over observed event count.}
\label{tab:precipcat}
\numtable{\begin{tabular}{lrrrr}
\toprule
Model & POD & FAR & CSI & Freq.\ bias \\
\midrule
Jua EPT-2.1 Europa & 0.28 & 0.62 & 0.19 & 0.74 \\
DWD ICON Global & 0.29 & 0.65 & 0.19 & 0.82 \\
Jua EPT-2 Reasoning & 0.23 & 0.56 & 0.18 & 0.53 \\
ECMWF IFS & 0.25 & 0.65 & 0.17 & 0.71 \\
NOAA GFS & 0.25 & 0.66 & 0.17 & 0.74 \\
Jua EPT-2 HRRR & 0.20 & 0.53 & 0.16 & 0.42 \\
Jua EPT-2e & 0.20 & 0.55 & 0.16 & 0.44 \\
ECMWF ENS (mean) & 0.15 & 0.50 & 0.13 & 0.29 \\
\bottomrule
\end{tabular}
}
\end{table}

\FloatBarrier
\subsection{Country-level regime profiles}
\label{app:country}

\begin{figure}[htbp]
\centering
\includegraphics[width=\textwidth]{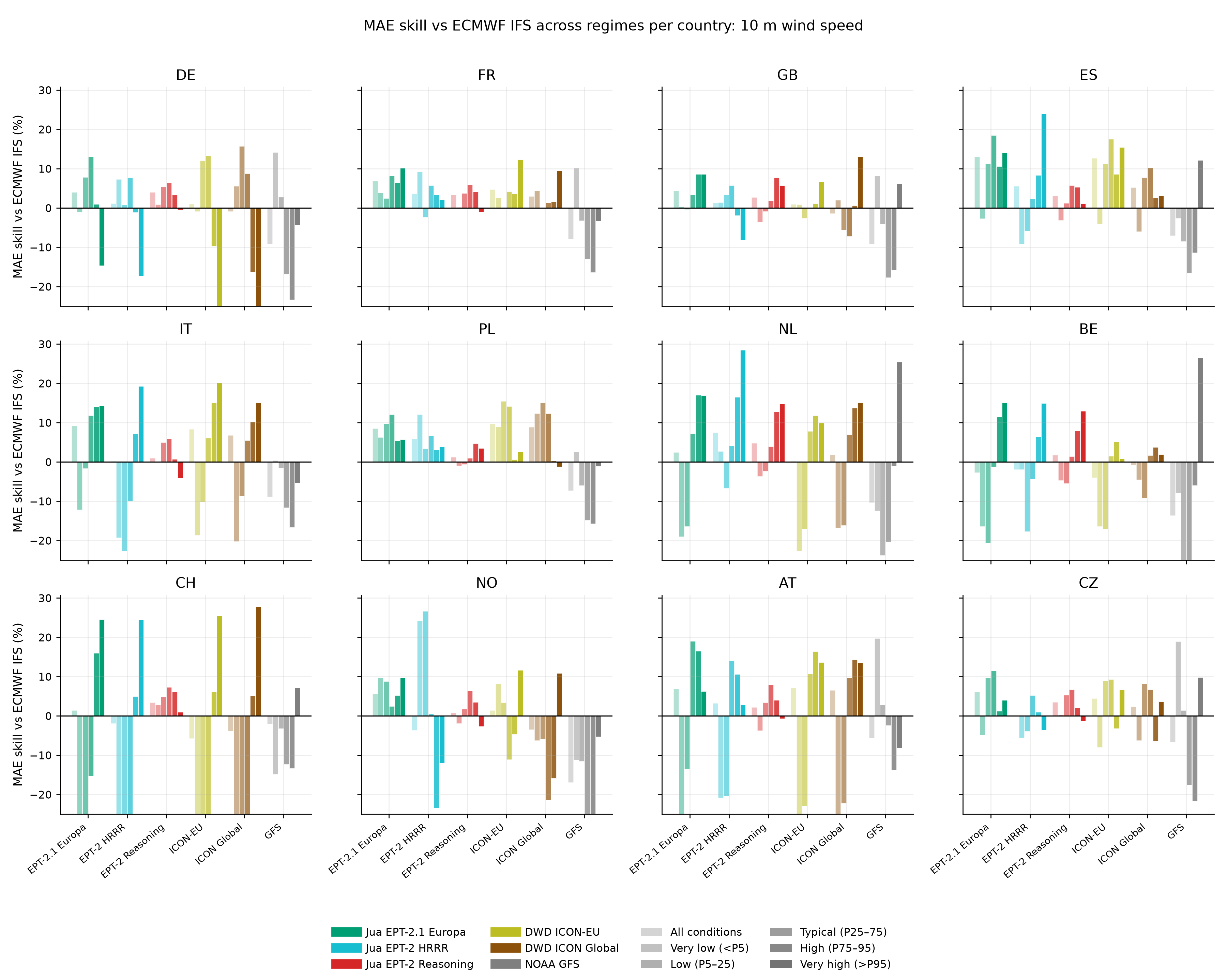}
\caption{Per-country MAE skill by ERA5 1991--2020 climatological regime,
10\,m wind, March--June~2026, pooled 1 to 48\,h (hourly). Mixed
regional/global model set.}
\label{fig:appbwind}
\end{figure}

\begin{figure}[htbp]
\centering
\includegraphics[width=\textwidth]{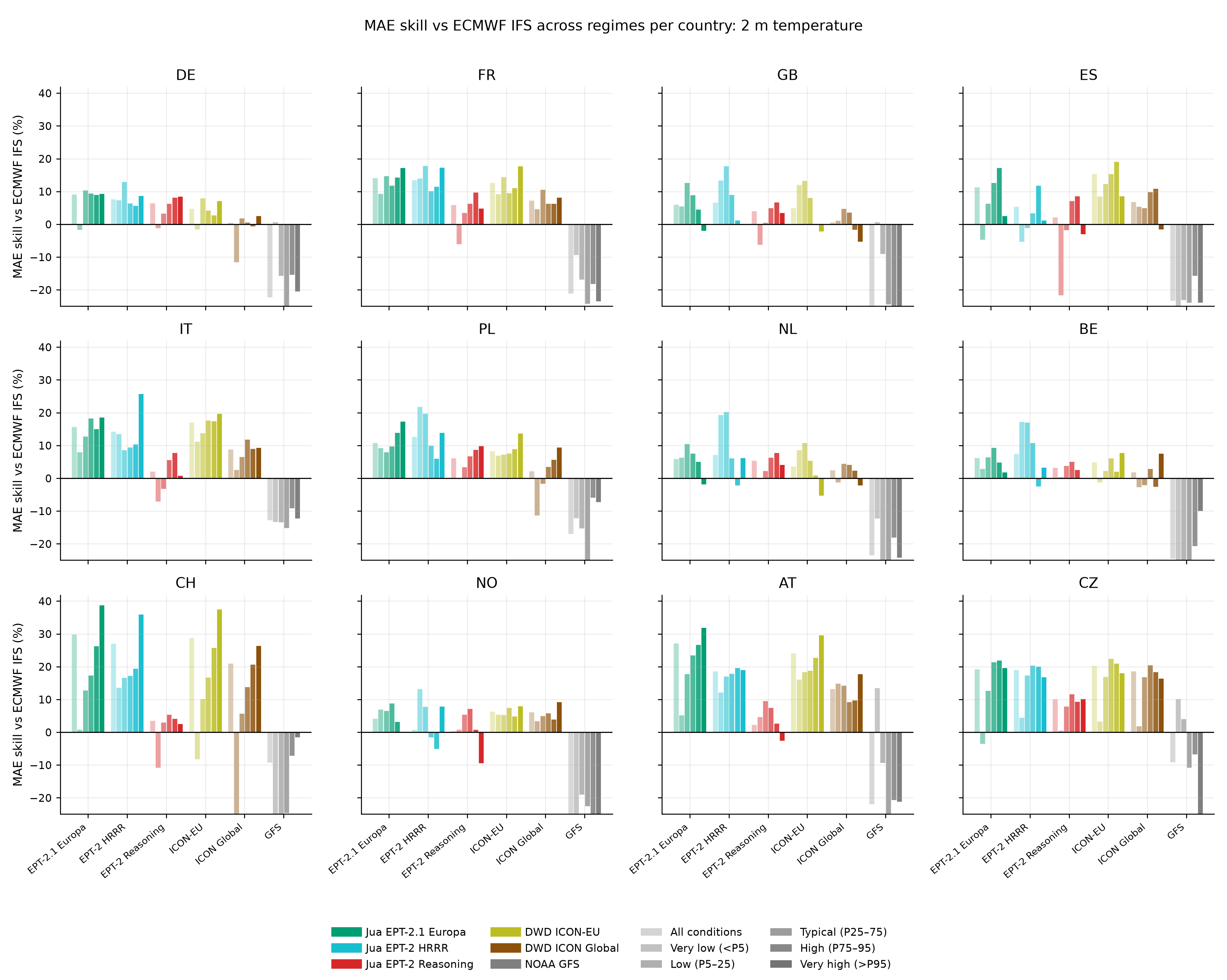}
\caption{As Figure~\ref{fig:appbwind}, for 2\,m temperature.}
\label{fig:appbtemp}
\end{figure}

\begin{figure}[htbp]
\centering
\includegraphics[width=\textwidth]{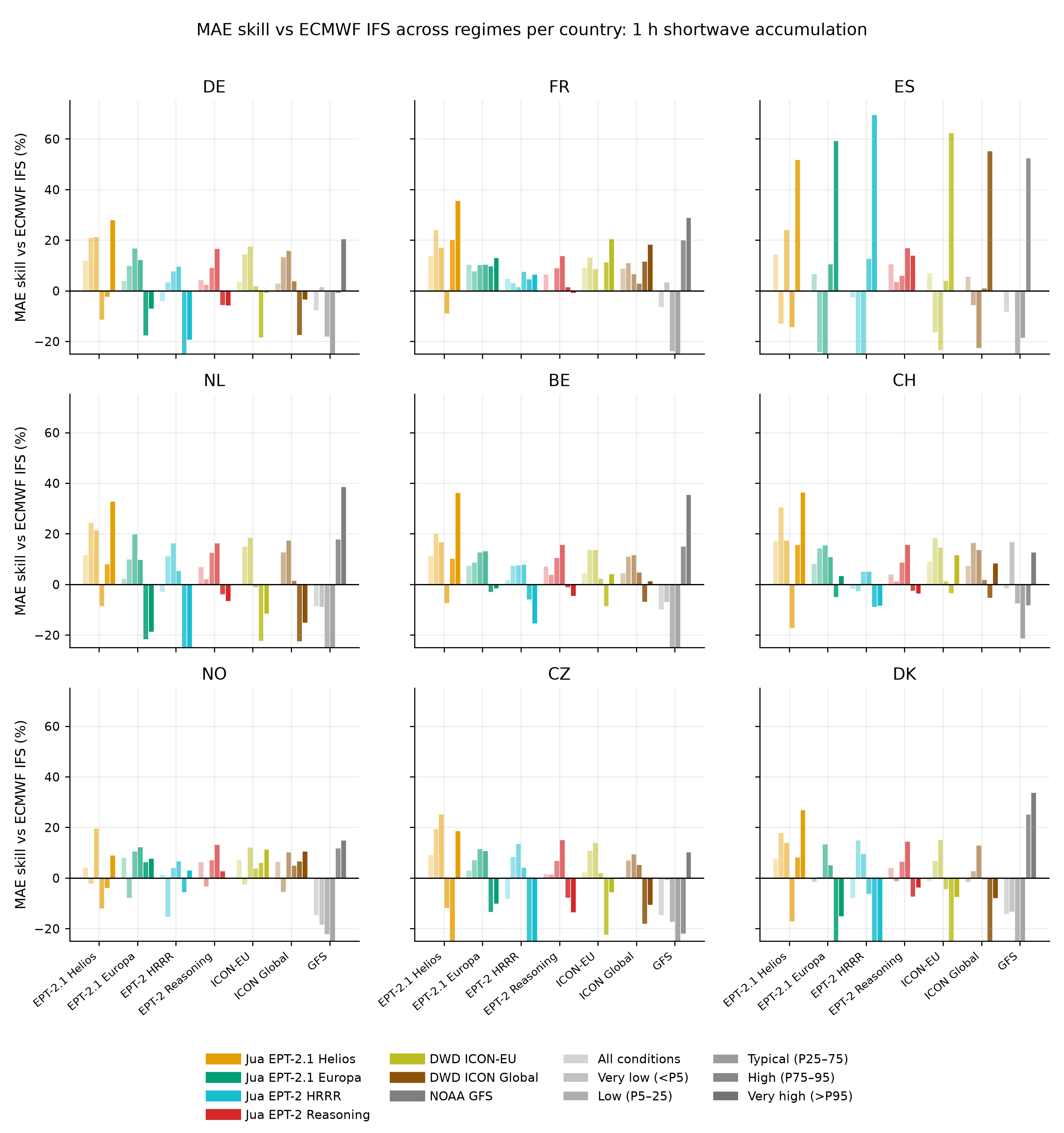}
\caption{As Figure~\ref{fig:appbwind}, for 1\,h shortwave, with EPT-2.1 Helios added.
GB, IT and PL have no solar station observations
(Section~\ref{sec:stations}).}
\label{fig:appbsolar}
\end{figure}

\begin{figure}[htbp]
\centering
\includegraphics[width=\textwidth]{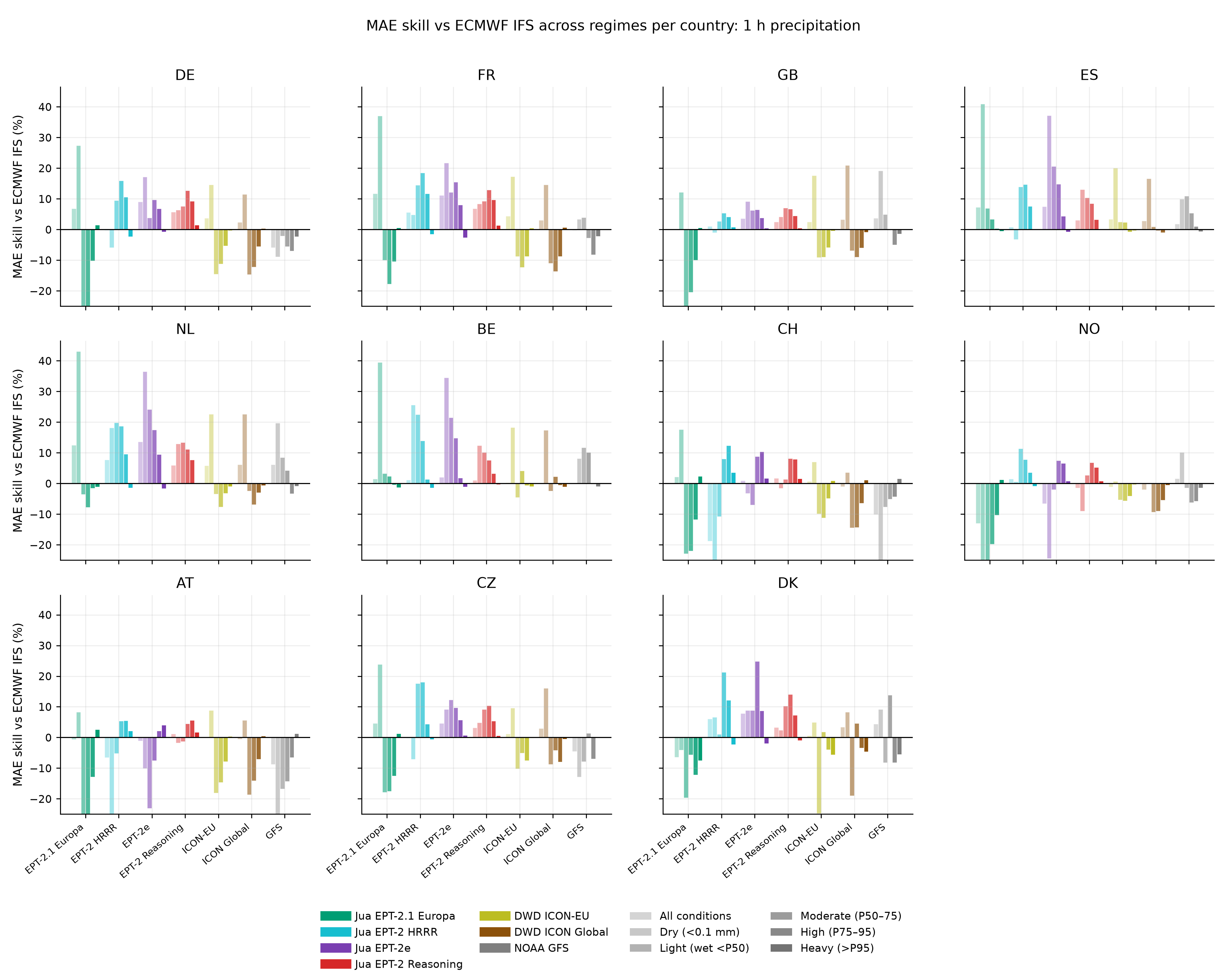}
\caption{Per-country MAE skill by precipitation regime for the mixed
regional/global model set. IT and PL have no gauges in the precipitation
panel.}
\label{fig:appbprecip}
\end{figure}

Figures~\ref{fig:appbwind}, \ref{fig:appbtemp}, \ref{fig:appbsolar}, and
\ref{fig:appbprecip} show how the March--June regime-skill patterns vary
across countries for the mixed regional/global comparison set. Countries
without sufficient matched observations for a variable are omitted.

\FloatBarrier
\subsection{Record-breaking temperature events and major windstorms}
\label{app:officialevents}

Table~\ref{tab:officialeventcatalog} lists the externally documented
temperature-record events and European windstorms used in the case-study
analysis. Event windows and affected countries were fixed from the source
catalogue before scoring. Every benchmark station in the source-reported
countries is included at 00, 06, 12, and 18\,UTC throughout the event
window. The analysis is therefore conditioned on documented events, rather
than on whether an individual station observation breaks a record.

For event \(e\) and lead
\(\ell\in\{6,12,\ldots,48\}\,\mathrm{h}\), let
\(\mathcal{C}_{e,\ell}\) denote the station--valid-time pairs \((s,t)\) in
the event footprint for which ECMWF IFS and every Track~A model included in
the comparison provide a forecast initialised at \(t-\ell\). Thus, each
model is evaluated against the same observations at each lead. For ensemble
models, \(f_m\) denotes the ensemble mean and \(\widetilde{f}_m\) the
forecast after the causal bias correction of Section~\ref{sec:debias}. We
compute
\begin{equation}
\begin{aligned}
\varepsilon_m(s,t,\ell)
&= \widetilde{f}_m(s,t\mid t-\ell)-y(s,t),\\
N
&= \sum_e \sum_\ell \lvert\mathcal{C}_{e,\ell}\rvert,\\
\operatorname{MAE}^{\mathrm{event}}_m
&= \frac{1}{N}\sum_e\sum_\ell
\sum_{(s,t)\in\mathcal{C}_{e,\ell}}
\lvert\varepsilon_m(s,t,\ell)\rvert,\\
S^{\mathrm{event}}_m
&= 100\left(
1-\frac{\operatorname{MAE}^{\mathrm{event}}_m}
{\operatorname{MAE}^{\mathrm{event}}_{\mathrm{IFS}}}
\right).
\end{aligned}
\label{eq:eventskill}
\end{equation}

The same observation may contribute at multiple leads because these are
distinct forecasts from different initialisation times. For uncertainty, we
delete one complete event at a time and report the jackknife standard error
\[
\sqrt{\frac{E-1}{E}\sum_e
\left(S^{\mathrm{event}}_{m,(-e)}
-\overline{S}^{\mathrm{event}}_{m,(-\cdot)}\right)^2},
\]
where \(E=5\) for temperature and \(E=7\) for wind.
Complete pooled values and event-level uncertainty are reported in
Table~\ref{tab:officialeventskill}.

\begin{table}[htbp]
\centering
\caption{Source-defined event catalogue. The accompanying reproduction
package records the source URLs and selection basis for every event.}
\label{tab:officialeventcatalog}
\numtable{\resizebox{\textwidth}{!}{
\begin{tabular}{@{}lllp{5.2cm}@{}}
\toprule
Variable & Event & Dates & Affected countries \\
\midrule
Temperature & Central European December record warmth & 2025-12-07--2025-12-09 & NL, CZ \\\addlinespace[0.2em]
Temperature & Late-February European record warmth & 2026-02-25--2026-02-27 & FR, DE, ES, CZ \\\addlinespace[0.2em]
Temperature & Early-April record warm spell & 2026-04-05--2026-04-08 & FR, GB \\\addlinespace[0.2em]
Temperature & Western European May heatwave & 2026-05-21--2026-05-30 & FR, GB \\\addlinespace[0.2em]
Temperature & Late-June European record heatwave & 2026-06-21--2026-06-28 & DE, FR, GB, ES, PL, NL, CH, AT, CZ, DK \\\addlinespace[0.2em]
Wind & North Sea autumn storm & 2025-09-15--2025-09-16 & NL \\\addlinespace[0.2em]
Wind & Storm Amy & 2025-10-03--2025-10-05 & GB, FR, BE, NL, NO, DK, PL \\\addlinespace[0.2em]
Wind & Storm Benjamin & 2025-10-23--2025-10-24 & FR, NL, CH, GB \\\addlinespace[0.2em]
Wind & Storm Goretti & 2026-01-08--2026-01-09 & GB, FR, DE, NL \\\addlinespace[0.2em]
Wind & Storm Nils & 2026-02-11--2026-02-13 & FR, ES \\\addlinespace[0.2em]
Wind & Storm Marlis & 2026-03-25--2026-03-26 & CH, DE \\\addlinespace[0.2em]
Wind & Storm Deborah & 2026-03-26--2026-03-27 & FR, IT \\
\bottomrule
\end{tabular}
}}
\end{table}

\begin{table}[htbp]
\centering
\caption{Event-case MAE skill (\%) relative to ECMWF IFS. Parentheses give
leave-one-event-out jackknife SE. Temperature uses five events and 231{,}065
common samples; wind uses seven events and 123{,}847 common samples.}
\label{tab:officialeventskill}
\numtable{\resizebox{0.78\textwidth}{!}{
\begin{tabular}{@{}lrr@{}}
\toprule
Model & Temperature events & Windstorms \\
\midrule
Jua EPT-2.1 Europa & +15.4\,(1.6) & +6.4\,(1.8) \\\addlinespace[0.25em]
Jua EPT-2 HRRR & +12.3\,(2.7) & +3.9\,(1.3) \\\addlinespace[0.25em]
Jua EPT-2 Reasoning & +6.5\,(0.6) & +3.3\,(0.5) \\\addlinespace[0.25em]
Jua EPT-2e & +4.3\,(1.1) & -0.3\,(0.8) \\\addlinespace[0.25em]
Microsoft Aurora & +2.5\,(0.9) & +0.7\,(0.9) \\\addlinespace[0.25em]
ECMWF AIFS & -3.4\,(3.9) & -6.0\,(1.3) \\\addlinespace[0.25em]
ECMWF AIFS ENS (mean) & +0.1\,(2.0) & -5.1\,(0.7) \\\addlinespace[0.25em]
ECMWF ENS (mean) & -0.9\,(0.1) & +1.7\,(0.7) \\\addlinespace[0.25em]
NOAA GFS & -25.4\,(7.5) & -12.6\,(1.8) \\\addlinespace[0.25em]
DWD ICON Global & +9.8\,(2.3) & +2.4\,(2.2) \\
\bottomrule
\end{tabular}
}}
\end{table}

\FloatBarrier
\end{document}